\documentclass[fleqn,usenatbib,useAMS]{mnras}
\usepackage[table,xcdraw]{xcolor}
\usepackage{newtxtext}
\usepackage[utf8]{inputenc}
\usepackage[english]{babel}
\usepackage[T1]{fontenc}
\usepackage{color}
\usepackage{deluxetable}
\usepackage{amssymb,amsmath}
\usepackage{rotating}
\usepackage{graphicx}
\usepackage{hyperref}
\hypersetup{colorlinks,citecolor=blue,filecolor=blue,linkcolor=blue,urlcolor=blue}
\usepackage{mathtools}
\usepackage{mathpazo}
\usepackage{graphicx}
\usepackage{footnote}
\usepackage{longtable}
\usepackage{soul}
\usepackage{caption}

\title[INOV of jetted vs. non-jetted RLNLSy1s]{Evidence of jet induced optical microvariability in radio-loud Narrow Line Seyfert 1 Galaxies}

\author[Ojha, VKJ, HC, VS ]{Vineet Ojha\thanks{E-mail: vineetojha@prl.res.in, vineetojhabhu@gmail.com}$^{1,~ 2}$, Vivek Kumar Jha$^{2,~3}$, Hum Chand$^{2,~4}$, Veeresh Singh$^{1}$   \\
  $^{1}$Physical Research Laboratory (PRL), Astronomy and Astrophysics Division, Ahmedabad, \it{380 009}; India \\
  $^{2}$Aryabhatta Research Institute of Observational Sciences (ARIES), Manora Peak, Nainital,
  \it{263002}; India\\
  $^{3}$Department of Physics, Deen Dayal Upadhyaya Gorakhpur University, Gorakhpur, \it {273009}; India\\
 $^{4}$Department of Physics \& Astronomical sciences, Central University of Himachal Pradesh, Dharamshala, \it{176215}; India\\}
 
\date{Accepted XXX. Received YYY; in original form ZZZ}
\pubyear{2022}

\begin{document}
\label{firstpage}
\pagerange{\pageref{firstpage}-- \pageref{lastpage}}
\maketitle
\begin{abstract}

To quantify the role of radio jets for Intra-Night Optical Variability (INOV) in Radio-Loud Narrow-Line Seyfert 1 (RLNLSy1) galaxies, we report the first systematic comparative INOV study of 23 RLNLSy1 galaxies, with 15 RLNLSy1s having confirmed detection of jets (jetted) and the remaining 8 RLNLSy1s having no detection of jets (non-jetted) based on their Very Long Baseline Array observations. We have monitored these two samples, respectively, in 37 and 16 sessions of a minimum 3-hour duration each. Based upon F$^{\eta}$-test at 99\% confidence level with a typical INOV amplitude ($\psi$) detection threshold of $>$ 3\%, we find the INOV duty cycles of 12\% for the sample of jetted RLNLSy1s, however, none of the sources showed INOV in the sample of non-jetted RLNLSy1s. Among the jetted RLNLSy1s, we find that the Duty Cycle (DC) for jetted $\gamma$-ray detected ($\gamma$-ray) RLNLSy1s is found to be 34\% in contrast to null INOV detection in the case of non-$\gamma$-ray RLNLSy1s. It suggests that instead of the mere presence of a jet, relativistic beaming plays a significant role for INOV in the case of low-luminous high accreting AGNs such as NLSy1s in which dilution of the AGN's non-thermal optical emission by the (much steadier) optical emission contributed by the nuclear accretion disc is quite likely. Our study of jetted $\gamma$-ray RLNLSy1s shows more frequent INOV detection for sources with higher apparent jet speed. Further, our results also suggest that among the NLSy1s, only jetted $\gamma$-ray RNLSy1 galaxies DC approaches blazar like DC.

\end{abstract}



\section{Introduction}
 \label{sec1.0}
Narrow-line Seyfert 1 (NLSy1) galaxies are a subclass of active galactic nuclei (AGN), emitting electromagnetic radiations from radio to gamma-ray wavebands. Although in the optical wavelengths, both permitted and forbidden emission lines are present in their spectra, the width of their broad component of Balmer emission lines is narrower than the population of general type-1 Seyfert galaxies, with the full width at half maximum of the broad component of Balmer emission line (FWHM(H${\beta}$)) being less than 2000 km s$^{-1}$~\citep{Osterbrock1985ApJ...297..166O, Goodrich1989ApJ...342..908G}. Other optical characteristics such as flux ratio of [O$_{III}]_{\lambda5007}/H\beta$ $<$ 3, and strong permitted Fe~{\sc ii} emission lines are used in addition to the criterion of FWHM(H${\beta}$) to characteristically define NLSy1 galaxies~\citep{Shuder-Osterbrock1981ApJ...250...55S}. Besides, these galaxies also display peculiar characteristics in other wavelength, especially in X-ray wavelength, such as strong soft X-ray excess below 2 keV~\citep[e.g.,][]{Brandt1997MNRAS.285L..25B, Vaughan1999MNRAS.309..113V, Vignali2004MNRAS.347..854V, Ojha2020ApJ...896...95O}, steep soft X-ray spectra~\citep[e.g.,][]{Boller1996A&A...305...53B, Wang1996A&A...309...81W, Grupe1998A&A...330...25G}, rapid X-ray (sometimes optical) flux variability~\citep[e.g.,][]{Leighly1999ApJS..125..297L, Komossa-Meerschweinchen2000A&A...354..411K, Miller2000NewAR..44..539M, Klimek2004ApJ...609...69K, Liu2010ApJ...715L.113L, Paliya2013MNRAS.428.2450P, Kshama2017MNRAS.466.2679K, Ojha2019MNRAS.483.3036O, Ojha2020MNRAS.493.3642O},  and blue-shifted emission line profiles~\citep[e.g.,][]{Zamanov2002ApJ...576L...9Z,Leighly2004ApJ...611..107L, Boroson2005AJ....130..381B,vivek2021}. Furthermore, NLSy1s are believed to be relatively young AGNs, and they represent an early phase of their evolution~\citep[e.g.,][]{Mathur2000MNRAS.314L..17M, Sulentic2000ApJ...536L...5S, Mathur2001NewA....6..321M, Fraix-Burnet2017FrASS...4....1F, Komossa2018rnls.confE..15K, Paliya2019JApA...40...39P}. Observationally, it is suggested that the majority of NLSy1s have relatively lower Super Massive Black Hole (SMBH) masses of 10$^{6}$ - 10$^{8}$ M$_{\sun}$~\citep{Grupe2004ApJ...606L..41G, Deo2006AJ....132..321D, Zhou2006ApJS..166..128Z, Peterson2011nlsg.confE..32P, Wang2014ApJ...793..108W, Rakshit2017ApJS..229...39R}, and higher accretion rates  $\lambda_{Edd} \sim$ 0.05 - 1.00, in contrast to luminous class of AGN such as quasars~\citep{Boroson1992ApJS...80..109B, Peterson2000ApJ...542..161P, Ojha2020ApJ...896...95O}.  However, relatively
lower SMBH mass is not uncontested since a systematic underestimation of their SMBH has been suggested~\citep{Decarli2008MNRAS.386L..15D, Marconi2008ApJ...678..693M, Calderone2013MNRAS.431..210C, Viswanath2019ApJ...881L..24V, Ojha2020ApJ...896...95O}. These highly accreting galaxies are generally hosted in spiral/disc galaxies~\citep{Deo2006AJ....132..321D, Ohta2007ApJS..169....1O, 2020MNRAS.492.1450O}, although in a few $\gamma$-ray detected NLSy1s, elliptical hosts have been suggested~\citep[hereafter $\gamma$-NLSy1s, ][]{D'Ammando2017MNRAS.469L..11D, D'Ammando2018MNRAS.478L..66D}.

Interestingly, although the NLSy1 exhibit both radio-quiet and radio-loud characteristics, defined by the radio parameter R$_{5 GHz}\equiv f_{5 GHz}/f_{4400\AA}$ with R$\leq$ 10 and $>$ 10 are being used to parameterized radio-quiet and radio-loud AGNs, respectively~\citep[e.g., see,][]{Stocke1992ApJ...396..487S, Visnovsky1992ApJ...391..560V, Kellermann1994AJ....108.1163K, Kellermann1989AJ.....98.1195K}, the population is dominated by radio-quiet objects~\citep{Kellermann2016ApJ...831..168K} and only a small fraction $\sim$ 7\% of NLSy1s are radio-loud~\citep[hereafter RLNLSy1s,][]{Komossa2006AJ....132..531K, Zhou2006ApJS..166..128Z, Rakshit2017ApJS..229...39R, Singh2018MNRAS.480.1796S}. This suggests that in a few of these galaxies, jets may be present, making them radio-loud~\citep{Zhou2003ApJ...584..147Z, Yuan2008ApJ...685..801Y}. Indeed, Very Long Baseline Array (VLBA) observations have discovered parsec-scale blazar-like radio jets in a few RLNLSy1s~\citep{Lister2013AJ....146..120L, Gu2015ApJS..221....3G, Lister2016AJ....152...12L}.

The existence of relativistic jets in such a subclass of AGN (although in a few of the sources) that has relatively higher accretion rates and lower black hole masses contradicts the general trend of the existence of relativistic jets in larger black hole masses and lower accretion rates~\citep{Urry2000ApJ...532..816U, Boroson2002ApJ...565...78B, 2002ApJ...564...86B, Urry2003ASPC..290....3U, Marscher2009arXiv0909.2576M, Chiaberge2011MNRAS.416..917C}, and also objects the theoretical paradigm of jet formation~\citep[e.g., ][]{2002ApJ...564...86B}. Hence, studying NLSy1s from the standpoint of jets is essential to understanding the physical processes that can launch relativistic jets in this subclass of AGN.\par   
 Nonetheless, despite a blazar-like double-humped spectral energy distribution (SED) of a few RLNLSy1s~\citep[e.g.,][]{Abdo2009ApJ...707L.142A, Paliya2013ApJ...768...52P, Paliya2019JApA...40...39P}, a minuscule fraction of RLNLSy1s, especially very radio-loud (R $>$ 100) RLNLSy1s exhibit interesting multi-wavelength characteristics such as compact radio cores, high brightness temperature, superluminal motion, flat radio and X-ray spectra, and rapid infrared and X-ray flux variability similar to blazar class of AGN~\citep{Boller1996A&A...305...53B, Grupe1998A&A...330...25G, Leighly1999ApJS..125..297L, Hayashida2000NewAR..44..419H, Komossa-Meerschweinchen2000A&A...354..411K, Yuan2008ApJ...685..801Y, Jiang2012ApJ...759L..31J, Orienti2012arXiv1205.0402O, Itoh2013ApJ...775L..26I, Yao2015MNRAS.454L..16Y, Berton2018A&A...614A..87B, Gabanyi2018rnls.confE..42G, Lister2018rnls.confE..22L}. All these characteristics give indirect evidence of the presence of jets in them. However $\gamma$-ray detection by {\it Fermi}-Large Area Telescope ({\it Fermi}-LAT)\footnote{https://heasarc.gsfc.nasa.gov/docs/heasarc/missions/fermi.html} from about two dozen RLNLSy1s gives conclusive evidence that $\gamma$-ray detected NLSy1s are capable of ejecting relativistic jets~\citep{Abdo2009ApJ...699..976A, Abdo2009ApJ...707..727A, Abdo2009ApJ...707L.142A, Foschini2010ASPC..427..243F, Foschini2011nlsg.confE..24F, D'Ammando2012MNRAS.426..317D, D'Ammando2015MNRAS.452..520D, Yao2015MNRAS.454L..16Y, Paliya2018ApJ...853L...2P, Yang2018MNRAS.477.5127Y, Yao2019MNRAS.487L..40Y}. \par

Variability of an AGN's optical flux from a few minutes to a day time scales is variously known as microvariability~\citep{Miller1989Natur.337..627M}, Intraday Variability~\citep[IDV,][]{Wagner1995ARA&A..33..163W} or Intra-Night Optical Variability~\citep[INOV,][]{Gopal-Krishna1993A&A...271...89G}. This alternative tool is also used to indirectly verify the presence or absence of jets in other sub-classes of AGN because of the well established observational fact that for radio-loud jet dominated sources such as blazars, both INOV amplitude ($\psi$) and the duty cycle (DC) are found to be distinctively high in comparison to non-blazars, including weakly polarised flat-radio-spectrum (i.e., radio-beamed) quasars~\citep{Goyal2013MNRAS.435.1300G, Gopal-Krishna2018BSRSL..87..281G}. Interestingly, such an indirect tool has been used for a decade in searching for the Doppler boosted optical jets in low luminous AGNs such as NLSy1s and weak emission line QSOs~\citep[e.g., see][]{Liu2010ApJ...715L.113L, Paliya2013MNRAS.428.2450P, Kumar2015MNRAS.448.1463K, Kumar2016MNRAS.461..666K, Kumar2017MNRAS.471..606K, Ojha2018BSRSL..87..387O, Ojha2021MNRAS.501.4110O}. However, this indirect evidence is based upon the observed high amplitude, and duty cycle of INOV as seen in blazars consisting of strongly Doppler boosted jets. More importantly, an INOV study comparing a subclass of AGN with jets and without jets has not yet been explored so far. Therefore, to establish stronger INOV amplitude ($\psi$) with high DC as evidence of the existence of jet in an AGN, we have carried out an INOV study with two sub-samples of RLNLSy1s with and without radio VLBA jets.\par

The general consensus regarding the radio structures of NLSy1s has been that they harbor sizes of less than 300 pc~\citep{Ulvestad1995AJ....109...81U, Lister2018rnls.confE..22L} and are generally compact sources with a steep spectrum~\citep{Foschini2011nlsg.confE..24F, Foschini2012nsgq.confE..10F, Berton2015A&A...578A..28B}. The appearance of radio structures in the radio observations of AGNs depends upon the resolution and sensitivity of the radio telescopes; therefore, non-detection of the radio jets in the radio images of RLNLSy1s may not necessarily imply that they do not have jets. Here, we have selected our sources (see Sect.~\ref{section_2.0}) based on their available observations with the radio telescopes, which mainly consist of VLBA observations. 

Furthermore, as pointed above, $\gamma$-ray detections in several RLNLSy1s suggest the presence of relativistic jets in them; therefore, a comprehensive INOV study of the jetted with $\gamma$-ray detected  RLNLSy1s (hereafter J-$\gamma$-RLNLSy1s) and the jetted without $\gamma$-ray detected  RLNLSy1s (hereafter J-RLNLSy1s) is essential to understand the nature of their variability and jets. Therefore, we have also discussed the INOV nature of J-$\gamma$-RLNLSy1s and J-RLNLSy1s sub-samples in the present work.\par
The layout of this paper is as follows. In Sect.~\ref{section_2.0}, we outline the sample selection procedure. Sect.~\ref{section_3.0} provides details of our intra-night optical monitoring and the data reduction procedure. The statistical analysis is presented in Sect.~\ref{sec4.0}, and our main results, followed by a brief discussion, are given in Sect.~\ref{sec 5.0}. In Sect.~\ref{sec6.0}, we summarize our main conclusions.

\begin{table*}
  \begin{minipage}{175mm} 
\caption{The present sample of 23 RLNLSy1s galaxies selected for INOV monitoring.}
\label{tab:source_info}
\begin{tabular}{rlrcccccl}
 \hline
 \multicolumn{1}{c}
 
   {SDSS Name,\footnote{\small The SDSS names of the sources with {\it Fermi}-LAT detection are suffixed with a ``$\blacktriangle$'' sign and the references~\citep{Abdo2009ApJ...699..976A, Abdo2009ApJ...707..727A, Foschini2010ASPC..427..243F};~\citep{Abdo2009ApJ...707L.142A};~\citep{Foschini2011nlsg.confE..24F, D'Ammando2012MNRAS.426..317D};~\citep{D'Ammando2015MNRAS.452..520D};~\citep{Liao2015arXiv151005584L};~\citep{Yao2015MNRAS.454L..16Y}; and~\citep{Ajello2020ApJ...892..105A} are for the sources, J094857.32$+$002225.6; J032441.20$+$341045.0 and J150506.48$+$032630.8; J084957.98$+$510829.0; J164442.53$+$261913.3; J130522.75$+$511640.2, J122222.99$+$041315.9, and J144318.60$+$472557.0 respectively.}} &  R-mag{\footnote{\small Taken from~\citet{Monet1998AAS...19312003M}.}}  & $z${\footnote{\small Emission-line redshifts are taken either from~\citet{Gu2015ApJS..221....3G} or from~\citet{Paliya2019ApJ...872..169P}.}}  &  $R_{1.4 GHz}${\footnote{\small $R_{1.4 GHz}\equiv f_{1.4 GHz}/f_{4400\AA}$ values are taken from~\citet{Gu2015ApJS..221....3G} for the sources marked with a `$^{\dagger}$', and are taken from~\citet{Ojha2020MNRAS.493.3642O} for the sources marked with an `$^{\star}$', except for J120014.08$-$004638.7 for which $R_{1.4 GHz}$ is estimated using its total flux density of 27.1 mJy at 1.4 GHz and k-corrected B-band optical flux density of 0.16 mJy~\citep{Doi2012ApJ...760...41D}.}} & Apparent jet speed{\footnote{\small The available jet speed of RLNLSy1s from literature which is as follows: for the jet speed of J032441.20$+$341045.0, J084957.98$+$510829.0, J094857.32$+$002225.6, J150506.48$+$032630.8 see~\citet[][]{Lister2019ApJ...874...43L}, and for J122222.99$+$041315.9, J164442.53$+$261913.3 see~\citealp[][]{Lister2016AJ....152...12L},~\citealp[and][]{Doi2012ApJ...760...41D}.}} & Optical Polarization{\footnote{\small The optical polarization values as reported in these papers: $^\alpha$\citet{Itoh2014PASJ...66..108I}; $^\beta$\citet{Ikejiri2011PASJ...63..639I}; $^\gamma$\citet{Angelakis2018arXiv180702382A}; $^\delta$\citet{Maune2014ApJ...794...93M}; $^\kappa$\citet{Leighly1999ApJS..125..297L}.}} & Radio Polarization{\footnote{\small The radio polarization values as reported in these papers: $^\psi$\citet{Neumann1994A&AS..106..303N}; $^\Lambda$\citet{Hodge2018ApJ...862..151H}; $^\tau$\citet{Homan2001ApJ...556..113H}. $^\star$fractional radio polarisation images presented in Fig-6 of~\citet{Gu2015ApJS..221....3G}.}} & log ($M_{BH}$){\footnote{\small Derived black hole masses of the present sample based upon single-epoch optical spectroscopy virial method were compiled from the literature. The references for the black hole mass are as follows: $^{\vee}$\citet{Zhou2007ApJ...658L..13Z}; $^{\diamond}$\citet{Yuan2008ApJ...685..801Y}; $^{\curlyvee}$\citet{Yao2015MNRAS.454L..16Y}; $^{\perp}$\citet{Rakshit2017ApJS..229...39R}; $^{\dotplus}$\citet{Wang2001A&A...377...52W}; $^{\zeta}$\citet{Greene2007ApJ...667..131G}.}} & Observing 
   \\

   & & & &   & & & \multicolumn{1}{c}{($M_{\sun}$)} & \multicolumn{1}{l}{freq.~(GHz)} \\\hline     
  \multicolumn{9}{c}{jetted NLSy1s}\\
 \hline 
J032441.20$+$341045.0$_{\blacktriangle}$     & 13.10 & 0.06 &   318$^{\star}$    & $9.1c\pm0.3c$ & 1-3\%$^{\alpha}$, 0.7-0.8\%$^{\beta}$, 1.2\%$^\gamma$ & 4\%$^\psi$,  0.2-1\%$^\Lambda$& 7.30$^{\vee}$ & 2.2/8.4 \\
J081432.12$+$560958.7    & 18.10 & 0.51 &  339$^{\dagger}$  & - & - & $\star$ & 8.00$^{\diamond}$& 4.9 \\
J084957.98$+$510829.0$_{\blacktriangle}$    & 17.79 & 0.58 & 4496$^{\star}$    & $6.6c\pm0.6c$ & 10\%$^{\delta}$, 10\%$^{\gamma}$ & 0.3-3\%$^\Lambda$, 3.3\%$^\tau$ & 7.59$^{\perp}$& 5.0/8.4/15.3 \\
J090227.20$+$044309.0    & 18.20 & 0.53 & 1047$^{\dagger}$  & - & - & $\star$   & 7.64$^{\perp}$ & 4.9 \\
J094857.32$+$002225.6$_{\blacktriangle}$    & 18.17 & 0.58 &  846$^{\star}$    & $9.7c\pm1.1c$ & 36\%$^{\alpha}$, 18.8\%$^{\beta}$, 2.4\%${^\gamma}$ & 0.2-3\%$^\Lambda$ & 7.50$^{\perp}$ & 22.2 \\
J095317.10$+$283601.5    & 18.60 & 0.66 &  513$^{\dagger}$  & - & - & -  & 7.73$^{\perp}$&  4.9 \\
J104732.78$+$472532.0    & 18.20 & 0.80 & 7413$^{\dagger}$  & - & - & - & 8.10$^{\diamond}$& 4.9 \\
J122222.99$+$041315.9$_{\blacktriangle}$    & 17.06 & 0.97 & 1534$^{\star}$   & $0.9c\pm0.3c$ & - & 0.2-3.3\%$^\Lambda$  & 8.30$^{\curlyvee}$& 15.4\\
J130522.75$+$511640.2$_{\blacktriangle}$    & 15.80 & 0.79 &  219$^{\dagger}$ & - & 1.0\%$^{\gamma}$ & $\star$  & 8.15$^{\perp}$ & 4.9  \\
J142114.05$+$282452.8    & 17.10 & 0.78 &  205$^{\dagger}$ & - & - & - & 7.72$^{\perp}$&  4.9 \\
J144318.60$+$472557.0$_{\blacktriangle}$    & 17.70 & 0.70 & 1175$^{\dagger}$ & - & - & $\star$  & 7.80$^{\diamond}$& 4.9  \\
J150506.48$+$032630.8$_{\blacktriangle}$    & 17.72 & 0.41 & 3364$^{\star}$   & $0.1c\pm0.2c$ & 4\%$^{\gamma}$ & 0.2-2.5\%$^\Lambda$ & 7.26$^{\perp}$ & 15.3\\
J154817.92$+$351128.0    & 18.40 & 0.48 &  692$^{\dagger}$  &-  & 2.1\%$^{\gamma}$ &  $\star$  &7.84$^{\perp}$& 4.9 \\
J164442.53$+$261913.3$_{\blacktriangle}$    & 16.60 & 0.14 &  447$^{\star}$  & $>1.0c$ & 2.2\%$^{\gamma}$ & - & 7.21$^{\perp}$ & 1.7 \\
J170330.38$+$454047.3    & 12.80 & 0.06 &  102$^{\star}$   & - & 3-5\%$^{\kappa}$ & - & 6.77$^{\dotplus}$& 1.7 \\

\hline
\multicolumn{9}{c}{non-jetted NLSy1s}\\
\hline
J085001.17$+$462600.5    & 18.40 & 0.52 &  170$^{\dagger}$ & - & - & - & 7.34$^{\perp}$& 4.9  \\
J103727.45$+$003635.6    & 19.10 & 0.60 &  457$^{\dagger}$ & - & - & - & 7.48$^{\perp}$& 4.9  \\
J111005.03$+$365336.2    & 19.00 & 0.63 &  933$^{\dagger}$ & - & - & - & 7.43$^{\perp}$ & 4.9  \\
J113824.53$+$365327.2    & 18.30 & 0.36 &  219$^{\dagger}$ & - & - & - & 7.29$^{\perp}$& 4.9  \\
J120014.08$-$004638.7    & 17.70 & 0.18 &  169           & - & - & -           & 7.40$^{\zeta}$ & 1.4\\
J124634.65$+$023809.1    & 17.50 & 0.36 &  277$^{\dagger}$ & - & - & - & 7.42$^{\perp}$ & 4.9  \\
J163323.59$+$471859.0    & 14.50 & 0.12 &  154$^{\star}$  & -  & 2.4\%$^{\gamma}$ & - & 6.70$^{\perp}$ & 1.7 \\
J163401.94$+$480940.2    & 19.10 & 0.49 &  204$^{\dagger}$ &-  & - & - & 7.56$^{\perp}$ &  4.9 \\
 \hline
\end{tabular}
 \end{minipage}
\end{table*}

\begin{table*}
 \begin{center}
  \caption{Details of system parameters of the telescopes and detectors used in the observations of 23 RLNLSy1s. }
  \label{tab:telescope_info}
  \begin{tabular}{c c c c c c c c c}
		\hline 
	Telescope (s) 	& No. of sessions$^{e}$	   & Detector (s)     &  \multicolumn{1}{l}{Field of view}    & Readout  & Gain        & Focal ratio & Pixel size & Plate scale\\
		                   &                  &(arcmin$^{2}$)     & noise     & (e$^-$ & of & of CCD     & of CCD\\
                                   &                  &                  &  (e$^-$)              &    /ADU)         &     telescope   & ($\mu$m) & ( $^{\prime\prime}$/pixel )   \\       
\hline
 1.04-m ST$^{a}$   &   2      & 4k$\times$4k     & 15.70$\times$15.70 & 3.0     &10.0 & f/13  & 15.0 & 0.23  \\
 1.30-m DFOT$^{b}$ & 47 & 2k$\times$2k     & 18.27$\times$18.27& 7.5     & 2.0 & f/4   & 13.5 & 0.53 \\
 3.60-m DOT$^{c}$   & 3   & 4k$\times$4k     &  6.52$\times$6.52 & 8.0$^\star$, 5.0$^\dagger$& 1.0$^\star$, 2.0$^\dagger$  & f/9   & 15.0 & 0.10  \\
 2.01-m HCT$^{d}$   & 1   & 2k$\times$2k     & 10.24$\times$10.24& 4.1     &2.20  & f/9   & 15.0 & 0.30  \\		
		\hline

 \multicolumn{8}{l}{$^{a}$Sampurnand Telescope (ST), $^{b}$Devasthal Fast Optical Telescope (DFOT), $^{c}$Devasthal Optical Telescope (DOT),}\\
 \multicolumn{8}{l}{$^{d}$Himalayan Chandra Telescope (HCT). $^{e}$ No. of intra-night sessions taken from the telescopes.}\\
             
 \multicolumn{8}{l}{$^\star$Readout noise and corresponding gain at readout speed of 1 MHz and `$^\dagger$' represents the same at readout speed of 500 kHz.}\\

  \end{tabular}
\end{center}

\end{table*}

\section{sample selection}
\label{section_2.0}
The bulk of our sample for intra-night monitoring is drawn from~\citet{Gu2015ApJS..221....3G} where they have reported good quality VLBA observations at 5 GHz for the 14 RLNLSy1 galaxies having a flux density above 10 mJy at 1.4 GHz and a radio-loudness parameter\footnote{R$_{1.4 GHz}$ is the ratio of the monochromatic rest-frame flux densities at 1.4 GHz and 4400\AA~\citep[see][]{Yuan2008ApJ...685..801Y}.} R$_{1.4 GHz}$ > 100.  Out of 14 RLNLSy1s, they confirm the presence of a jet in 7 of the sources (hereafter, "jetted") based on the detection of a core-jet structure at 5 GHz, and the remaining 7 sources are termed as "non-jetted" based on the detection of a compact core only. However, in reference to the jetted nature, it may be noted that VLBA radio images at typical resolution could generally resolve parsec scale core jet structures~\citep{Lister2018rnls.confE..22L} which may not be relativistically beamed. Hence some of the sources could be steep spectrum sources as seen in mini-radio galaxies \citep[see][]{Gu2015ApJS..221....3G}. We also note that the "non-jetted" source J144318.56$+$472556.7 has a $\sim$ 15-mas long quasi-linear radio component resolved into seven components along the south-west direction, in addition to diffuse emission extending up to $\sim$ 30 mas~\citep[see figure 12 of][]{Gu2015ApJS..221....3G}. Therefore, we have included this source in our jetted subsample. We next expanded this sample by including another 9 sources from the VLBA literature, which also satisfy the above twin criteria (i.e., $f_{\nu~1.4 GHz}\geq$ 10 mJy and R$_{1.4 GHz} > 100$). The jetted (or non-jetted) classification of the 8 sources out of 9 is possible using their published VLBA observations. Thus, three of these 8 sources, viz., J163323.59$+$471859.0, J164442.53$+$261913.3 and J170330.38$+$454047.3 are taken from the~\citet{Doi2011ApJ...738..126D} and the remaining 5 sources, viz., J032441.20$+$341045.0, J084957.98$+$510829.0, J094857.32$+$002225.6, J122222.99$+$041315.9 and J150506.48$+$032630.8 are taken from~\citet{Zhou2007ApJ...658L..13Z},~\citet{D'Ammando2012MNRAS.426..317D},~\citet{Giroletti2011A&A...528L..11G},~\citet{Lister2016AJ....152...12L} and~\citet{Orienti2012arXiv1205.0402O}, respectively. The frequencies at which observations of 23 RLNLSy1s were carried out are tabulated in the last column of Table~\ref{tab:source_info}. However, for the NLSy1 J120014.08$-$004638.7,~\citet{Doi2012ApJ...760...41D} have confirmed its lobe-dominated nature based upon its Very Large Array (VLA) 1.4 GHz FIRST images. Note that this source is also not in the latest sample of Monitoring Of Jets in Active galactic nuclei with VLBA Experiments (MOJAVE) XVII program~\citep[see][]{Lister2019ApJ...874...43L}. Therefore, we have included this source in our non-jetted set. Based on the published VLBA observations of 8 sources, 7 sources have a confirmed jet, and one source falls in the non-jetted category. Thus, overall, our sample consists of 23 RLNLSy1s, including 15 of which are jetted, and the remaining 8 RLNLSy1s are non-jetted. Table~\ref{tab:source_info} summarizes the basic properties of our sample. The SDSS names of the sources with {\it Fermi}-LAT detection are suffixed with a "$\blacktriangle$" sign, and the references are given in the footnote "b" to  Table~\ref{tab:source_info}. \par

\section{Observations and Data Reduction}
\label{section_3.0}

\subsection{Photometric monitoring observations}
Intra-night monitoring of all 23 RLNLSy1s of our sample was performed in the broad-band Johnson-Cousin filter R due to the optimum response of the used CCDs in this filter. Four telescopes namely, the 1.04 meter (m) Sampurnanand telescope~\citep[ST,][]{Sagar1999Csi...77...77.643}, 1.30-m Devasthal Fast Optical Telescope~\citep[DFOT,][]{Sagar2010ASInC...1..203S}, 3.60-m Devasthal Optical Telescope~\citep[DOT,][]{Sagar2012SPIE.8444E..1TS} and 2.01-m Himalayan Chandra Telescope~\citep[HCT,][]{Prabhu2010ASInC...1..193P} were used for the intra-night monitoring of the present sample. Out of these four telescopes, the 1.04-m ST is located at Nainital, while the 1.30-m DFOT and the 3.60-m DOT are located at Devasthal near Nainital, and all the three are managed by the Aryabhatta Research Institute of Observational Sciences (ARIES). The fourth telescope, the 2.01-m Himalayan Chandra Telescope (HCT),  is located in Ladakh and operated by the Indian Institute of Astrophysics (IIA), Bangalore, India. All the four telescopes are equipped with Ritchey-Chretien (RC) optics and were read out 1 MHz rate during our observations, except for the 3.60-m DOT, which was read out additionally at 500 kHz. The monitoring sessions lasted between $\sim$ 3.0 and $\sim$ 5.5 hours (median 3.75 hrs). Our sources were observed in 4$\times$4 binning mode with 1.04-m ST and the same with the 3.6-m DOT on 2017.04.11. A 2$\times$2 binning mode was adopted for the remaining sessions with the DOT. No binning was done for the DFOT and the HCT telescopes observations. The basic parameters of the four telescopes, the number of monitoring sessions, and the CCDs used in the present observations are listed in Table~\ref{tab:telescope_info}. 
In order to improve the INOV statistics, at least two intra-night sessions were managed for each of our 23 RLNLSy1s. In this work, 1.04-3.60m class telescopes have been used; therefore, depending {on the brightness} of the target NLSy1, telescope used, moon illumination, and sky condition, a typical exposure time for each science frame were set between 4 and 15 minutes in order to get a reasonable SNR. The median seeing (FWHM of the point spread function (PSF)) for the sessions typically ranged between $\sim$ 1 - 3 arcsec, except for a single session dated 2019.03.25 when seeing became considerably poorer (see Fig.~\ref{fig:lurve 5}).

\subsection{Data reduction}
\label{sec3.2}

For each night, sky flat-field images were taken during dusk and dawn, and at least three bias frames were taken. The dark frames were not taken during our observations due to the relatively low temperature of the CCD detectors used, which were cooled either using liquid nitrogen (to about $-120^\circ$C) or using thermoelectrical cooling (to about $-90^\circ$C in case of 1.3-m DFOT). The standard routines within the {\sc IRAF}\footnote{Image Reduction and Analysis Facility (http://iraf.noao.edu/)} software package were followed for preliminary processing of the observed frames. Aperture photometry~\citep{1987PASP...99..191S, 1992ASPC...25..297S} was selected in this work for extracting the instrumental magnitudes of the targets and the comparison stars recorded in the CCD frames, using DAOPHOT II algorithm\footnote{Dominion Astrophysical Observatory Photometry (http://www.astro.wisc.edu/sirtf/daophot2.pdf)} due to less crowded fields of the monitored NLSy1s. The prime parameter in the aperture photometry is the size of the optimal aperture, which is used to estimate the instrumental magnitude and the corresponding signal-to-noise ratio (SNR) of the individual photometric data points recorded in each CCD frame. As emphasized in~\citet{Howell1989PASP..101..616H}, the SNR of a target recorded in a CCD is maximized for the aperture radius $\sim$ PSF.
However, as suggested by~\citet{Cellone2000AJ....119.1534C} when the underlying host galaxy significantly contributes to the total optical flux, its contribution to the aperture photometry can vary significantly due to PSF variation, mimicking INOV.  Possibility of such spurious INOV can be significant for the lower redshift NLSy1s in our sample, particularly, J032441.20$+$341045.0 (z = 0.06), J164442.53$+$261913.3 (z = 0.14), J170330.38$+$454047.3 (z = 0.06), J120014.08$-$004638.7 (z = 0.18) and J163323.59$+$471859.0 (z = 0.12) (Table~\ref{tab:source_info}). This issue is further addressed in Sect.~\ref{sec 5.0}. Nonetheless, bearing the above in mind, we have chosen an aperture radius equal to 2$\times$FWHM for our final analysis, as already elaborated in~\citet{Ojha2021MNRAS.501.4110O}.\par

Using the instrumental magnitudes extracted from the aperture photometry, DLCs of each NLSy1s were derived for each session relative to a minimum of two (steady) comparison stars that were chosen based on their closeness to the monitored target NLSy1, both in position and brightness, as recorded in the CCD frames. The importance of these procedures for genuine INOV detection has been highlighted by~\citet{Howell1988AJ.....95..247H} and further focused in~\citet{Cellone2007MNRAS.374..357C}. In the case of 12 targets, we could identify at least a comparison star within $\sim$ 1 instrumental magnitude to the target NLSy1. The median magnitude offsets ($\Delta m_{R}$) for the remaining 10 targets were also not significant and within $\sim$ 1.5-mag, except for a source, viz., J103727.45$+$003635.6 for which $\Delta m_{R}$ was found to be 1.74 (see Figs 2-6). Table~\ref{tab_jnj_comp_star} lists coordinates together with some other parameters of the steady comparison stars used for all the sessions. It has been shown in~\citet{Ojha2021MNRAS.501.4110O} that the color differences of such orders can be safely discounted while analyzing the variability of the DLCs.

\onecolumn
\begin{center}
\LTcapwidth=\textwidth  
\begin{longtable}{ccc ccc c}
\caption{Basic parameters of the comparison stars along with their observation dates used in this study for the 23 RLNLSy1 galaxies.}
  \label{tab_jnj_comp_star} \\

  \hline \multicolumn{1}{c}{{Target RLNLSy1s and}} & \multicolumn{1}{c}{{Date(s) of monitoring}} & \multicolumn{1}{c}{{R.A.(J2000)}}&\multicolumn{1}{c}{{Dec.(J2000)}} & \multicolumn{1}{c}{{\it g}} & \multicolumn{1}{c}{{\it r}} & \multicolumn{1}{c}{{\it g-r}}\\
\multicolumn{1}{c}{the comparison stars}          &      &   \multicolumn{1}{c}{(h m s)}       &\multicolumn{1}{c}{($^\circ$ $^\prime$ $^{\prime\prime}$)}   & \multicolumn{1}{c}{(mag)}   & \multicolumn{1}{c}{(mag)}   & \multicolumn{1}{c}{(mag)}     \\
 \multicolumn{1}{c}{(1)}      & \multicolumn{1}{c}{(2)}        & \multicolumn{1}{c}{(3)}           & \multicolumn{1}{c}{(4)}                              & \multicolumn{1}{c}{(5)}   & \multicolumn{1}{c}{(6)}   &  \multicolumn{1}{c}{(7)}     \\ \hline 
\endfirsthead

\multicolumn{7}{l}%
{{\tablename\ \thetable{} -- continued..}} \\
\hline \multicolumn{1}{c}{{Target AGN and}} & \multicolumn{1}{c}{{Date(s) of monitoring}} & \multicolumn{1}{c}{{R.A.(J2000)}}&\multicolumn{1}{c}{{Dec.(J2000)}} & \multicolumn{1}{c}{{\it g}} & \multicolumn{1}{c}{{\it r}} & \multicolumn{1}{c}{{\it g-r}}\\
\multicolumn{1}{c}{the comparison stars}          &      &   \multicolumn{1}{c}{(h m s)}       &\multicolumn{1}{c}{($^\circ$ $^\prime$ $^{\prime\prime}$)}   & \multicolumn{1}{c}{(mag)}   & \multicolumn{1}{c}{(mag)}   & \multicolumn{1}{c}{(mag)}     \\
 \multicolumn{1}{c}{(1)}      & \multicolumn{1}{c}{(2)}        & \multicolumn{1}{c}{(3)}           & \multicolumn{1}{c}{(4)}                              & \multicolumn{1}{c}{(5)}   & \multicolumn{1}{c}{(6)}   &  \multicolumn{1}{c}{(7)}     \\\hline 
\endhead

\hline \multicolumn{7}{r}{{Continued to next page}} \\
\endfoot
\endlastfoot
                                            \multicolumn{7}{c}{jetted NLSy1s}\\
\hline
J032441.20$+$341045.0 & 2016 Nov. 22, 23; Dec. 02; 2017 Jan. 03, 04& 03 24 41.20  &$+$34 10 45.00 & 14.50 & 13.70& 0.80$^{*}$   \\
S1                    & 2016 Nov. 22, Dec. 02                      & 03 24 53.68  &$+$34 12 45.62 & 15.60 & 14.40& 1.20$^{*}$    \\
S2                    & 2016 Nov. 22, Dec. 02                      & 03 24 53.55  &$+$34 11 16.58 & 16.20 & 14.40& 1.80$^{*}$    \\
S3              & 2016 Nov. 23                       & 03 24 10.92 &  $+$34 15 01.90 &  16.30 & 15.10 &  1.20$^{*}$    \\
{ S4}              &{  2016 Nov. 23}                         &{  03 24 14.04}  &{ $+$34 18 20.10} &{  15.80} &{  15.00}&{  1.00$^{*}$}    \\
{ S5}              &{  2017 Jan. 03}                         &{  03 24 14.92}  &{ $+$34 15 21.20} &{  15.90} &{  15.10}&{  0.80$^{*}$}    \\
{ S6}              &{  2017 Jan. 03}                         &{  03 24 08.44}  &{ $+$34 08 15.80} &{  15.80} &{  14.20}&{  1.60$^{*}$}    \\
{ S7}              &{  2017 Jan. 04}                         &{  03 24 38.14}  &{ $+$34 13 53.40} &{  15.90} &{  15.20}&{  0.70$^{*}$}    \\
{ S8}              &{  2017 Jan. 04}                         &{  03 24 14.08}  &{ $+$34 16 48.60} &15.40 & 14.80 & 0.60$^{*}$   \\
J081432.12$+$560958.7 & 2017 Jan. 03; Nov. 20        & 08 14 32.12  &$+$56 09 58.69 & 18.06 & 18.11 & \hspace{-0.35 cm}$-$0.05 \\ 
S1                    & 2017 Jan. 03                 & 08 14 02.78  &$+$56 11 12.07 & 19.14 & 18.12 & 1.02      \\ 
S2                    & 2017 Jan. 03                 & 08 14 53.54  &$+$56 10 14.14 & 17.96 & 17.35 & \hspace{-0.35 cm}$-$0.61 \\
S3                    & 2017 Nov. 20                 & 08 13 39.62  &$+$56 17 55.59 & 19.05 & 17.71 & 1.34 \\
S4                    & 2017 Nov. 20                 & 08 14 19.58  &$+$56 06 24.04 & 19.33 & 17.88 & 1.45 \\
J084957.98$+$510829.0 & 2017 Dec. 13, 2019 April 08  & 08 49 57.98  &$+$51 08 29.04 & 18.92 & 18.28 & 0.64\\
S1                    &                              & 08 50 12.62  &$+$51 08 08.03 & 19.45 & 18.06 & 1.39  \\
S2                    &                              & 08 50 03.07  &$+$51 09 12.23 & 17.82 & 17.09 & 0.73  \\
J090227.20$+$044309.0 & 2017 Feb. 22; Dec. 14        & 09 02 27.20  &$+$04 43 09.00 & 18.96 & 18.63 & 0.33 \\
S1                    & 2017 Feb. 22                 & 09 02 01.94  &$+$04 37 32.90 & 19.10 & 17.74 & 1.36  \\ 
S2                    & 2017 Feb. 22                 & 09 02 23.40  &$+$04 35 44.57 & 18.70 & 17.31 & 1.39     \\
S3                    & 2017 Dec. 14                 & 09 03 04.11  &$+$04 48 19.65 & 18.85 & 17.95 & 0.90    \\
S4                    & 2017 Dec. 14                 & 09 03 07.29  &$+$04 38 57.34 & 18.92 & 17.93 & 0.99    \\
J094857.32$+$002225.6 & 2016 Dec. 02; 2017 Dec. 21   & 09 48 57.32  &$+$00 22 25.56 & 18.59 & 18.43 & 0.16    \\
S1                    &                              & 09 48 36.95  &$+$00 24 22.55 & 17.69 & 17.28 & 0.41    \\
S2                    &                              & 09 48 37.47  &$+$00 20 37.02 & 17.79 & 16.70 & 1.09    \\
J095317.10$+$283601.5 & 2017 March 04; 2018 March 23 & 09 53 17.10  &$+$28 36 01.48 & 18.99 & 18.97 & 0.02    \\
S1                    & 2017 March 04                & 09 52 48.09  &$+$28 29 53.69 & 18.31 & 17.45 & 0.86     \\ 
S2                    & 2017 March 04; 2020 November 21 & 09 53 07.49  &$+$28 37 17.10 & 18.46 & 17.32 & 1.14     \\
S3                    & 2020 November 21             & 09 53 21.03  &$+$28 34 57.36 & 20.41 & 18.90 & 1.51   \\ 
J104732.78$+$472532.0 & 2017 April 11; 2018 March 12 & 10 47 32.78  &$+$47 25 32.02 & 18.97 & 18.76 & 0.21  \\
S1                    & 2017 April 11                & 10 47 16.50  &$+$47 24 47.24 & 18.68 & 17.98 & 0.70      \\
S2                    & 2017 April 11; 2018 March 12 & 10 47 27.51  &$+$47 27 58.94 & 18.79 & 17.87 & 0.92     \\
S3                    & 2018 March 12                & 10 48 16.44  &$+$47 22 42.22 & 18.78 & 17.45 & 1.33    \\
J122222.99$+$041315.9 & 2017 Jan. 03, 04; Feb. 21, 22; March 04, 24 & 12 22 22.99  &$+$04 13 15.95  & 17.02 & 16.80& 0.22  \\
S1                    &                                             & 12 22 34.02  &$+$04 13 21.57  & 18.63 & 17.19& 1.44      \\
S2                    &                                             & 12 21 56.12  &$+$04 15 15.19  & 17.22 & 16.78& 0.44    \\
J130522.75$+$511640.2 & 2017 April 04; 2019 April 25 & 13 05 22.74  &$+$51 16 40.26 & 17.29 & 17.10 & 0.19  \\
S1                    & 2017 April 04                & 13 06 16.16  &$+$51 19 03.67 & 16.96 & 15.92 & 1.04      \\
S2                    & 2017 April 04; 2019 April 25 & 13 05 57.57  &$+$51 11 00.97 & 16.35 & 15.26 & 1.09      \\
S3                    & 2019 April 25                & 13 05 44.25  &$+$51 07 35.85 & 17.88 & 16.42 & 1.46      \\
J142114.05$+$282452.8 & 2018  May  10; 2019  May  27 & 14 21 14.05  &$+$28 24 52.78 & 17.73 & 17.74 & \hspace{-0.25 cm}$-$0.01 \\ 
S1                    & 2018  May  10                & 14 20 33.73  &$+$28 31 10.45 & 18.50 & 17.11 & 1.39   \\ 
S2                    & 2018  May  10; 2019  May  27 & 14 21 08.78  &$+$28 24 04.99 & 16.16 & 16.21 & \hspace{-0.25 cm}$-$0.05  \\
S3                    & 2019  May  27                & 14 21 24.36  &$+$28 27 16.52 & 16.82 & 16.45 & 0.37 \\
J144318.56$+$472556.7 & 2018 March 11, 23            & 14 43 18.56  &$+$47 25 56.74 & 18.14 & 18.17 & \hspace{-0.25 cm}$-$0.03 \\
S1                    &                              & 14 43 37.14  &$+$47 23 03.03 & 17.51 & 16.82 & 0.69      \\
S2                    &                              & 14 43 19.05  &$+$47 19 00.98 & 18.03 & 16.75 & 1.28                          \\
J150506.48$+$032630.8 & 2017 March 25; 2018 April 12 & 15 05 06.48  &$+$03 26 30.84 & 18.64 & 18.22 & 0.42    \\
S1                    &                              & 15 05 32.05  &$+$03 28 36.13 & 18.13 & 17.64 & 0.49    \\
S2                    &                              & 15 05 14.52  &$+$03 24 56.17 & 17.51 & 17.14 & 0.37    \\
J154817.92$+$351128.0 & 2018  May  17; 2019  May  08 & 15 48 17.92  &$+$35 11 28.00 & 18.03 & 18.03 & 0.00    \\   
S1                    &                              & 15 47 57.43  &$+$35 14 05.24 & 18.31 & 17.50 & 0.81      \\ 
S2                    &                              & 15 48 02.20  &$+$35 13 56.16 & 17.37 & 16.98 & 0.39        \\
J164442.53$+$261913.3 & 2017 April 03; 2019 April 26 & 16 44 42.53  &$+$26 19 13.31 & 18.03 & 17.61 & 0.42    \\
S1                    &                              & 16 45 20.03  &$+$26 20 54.55 & 16.56 & 15.89 & 0.67    \\
S2                    &                              & 16 44 34.40  &$+$26 15 30.27 & 16.28 & 15.80 & 0.48    \\
J170330.38$+$454047.3 & 2017 June  03; 2019 March 25 & 17 03 30.38  &$+$45 40 47.27 & 16.12 & 15.41 & 0.71    \\
S1                    &                              & 17 04 02.02  &$+$45 42 16.56 & 15.02 & 14.39 & 0.63      \\ 
S2                    &                              & 17 04 34.88  &$+$45 40 08.65 & 15.00 & 13.91 & 1.09        \\\hline


                                         \multicolumn{7}{c}{non-jetted NLSy1s}\\
\hline
J085001.17$+$462600.5 & 2017 Jan. 01; Dec. 15        & 08 50 01.17  &$+$46 26 00.50 & 19.12 & 18.82 & 0.30     \\
S1                    &                              & 08 50 17.69  &$+$46 20 42.71 & 18.11 & 17.85 & 0.26        \\ 
S2                    &                              & 08 49 48.29  &$+$46 21 11.81 & 18.72 & 17.63 & 1.09       \\
J103727.45$+$003635.6 & 2018 March 11, 22            & 10 37 27.45  &$+$00 36 35.60 & 19.57 & 19.21 & 0.36     \\
S1                    & 2018 March 11                & 10 37 39.63  &$+$00 38 26.16 & 18.90 & 17.69 & 1.21       \\ 
S2                    & 2018 March 11                & 10 37 28.03  &$+$00 37 59.88 & 18.97 & 17.52 & 1.45        \\
S3                    & 2021 April 08                & 10 36 50.96  &$+$00 41 25.26 & 18.76 & 17.35 & 1.41       \\ 
S4                    & 2021 April 08                & 10 37 38.72  &$+$00 40 28.00 & 17.26 & 16.82 & 0.44        \\
J111005.03$+$365336.2 & 2018 March 23; 2019 Jan. 13  & 11 10 05.03  &$+$36 53 36.22 & 20.60 & 20.49 & 0.11     \\
S1                    &                              & 11 10 08.50  &$+$36 50 59.03 & 19.14 & 18.86 & 0.28       \\ 
S2                    &                              & 11 10 10.76  &$+$36 55 26.97 & 19.74 & 18.45 & 1.29        \\
J113824.53$+$365327.2 & 2017 April 17; 2018 March 23 & 11 38 24.53  &$+$36 53 27.18 & 19.55 & 18.79 & 0.76     \\
S1                    &                              & 11 38 25.03  &$+$36 54 44.02 & 18.90 & 17.52 & 1.38       \\ 
S2                    &                              & 11 37 56.81  &$+$36 52 35.56 & 17.91 & 17.41 & 0.50        \\
J120014.08$-$004638.7 & 2018 March 12, May 11        & 12 00 14.08  &$-$00 46 38.74 & 18.51 & 17.81 & 0.70     \\
S1                    &                              & 12 00 12.63  &$-$00 46 07.14 & 17.19 & 16.68 & 0.51       \\ 
S2                    &                              & 12 00 25.99  &$-$00 51 45.21 & 16.73 & 16.37 & 0.36         \\
J124634.65$+$023809.1 & 2017 April 03; 2018 April 12 & 12 46 34.65  &$+$02 38 09.06 & 18.35 & 18.18 & 0.17     \\
S1                    & 2017 April 03; 2018 April 12 & 12 47 00.55  &$+$02 37 31.37 & 17.91 & 16.89 & 1.02       \\ 
S2                    & 2017 April 03                & 12 47 05.32  &$+$02 39 06.75 & 16.94 & 16.62 & 0.32       \\
S3                    & 2018 April 12                & 12 46 49.50  &$+$02 37 11.64 & 17.24 & 16.77 & 0.47       \\
J163323.59$+$471859.0 & 2017  May  20; 2019 March 20 & 16 33 23.59  &$+$47 18 59.04 & 17.25 & 16.95 & 0.30     \\
S1                    &                              & 16 32 59.26  &$+$47 26 05.45 & 15.57 & 15.18 & 0.39        \\ 
S2                    &                              & 16 32 56.00  &$+$47 21 01.26 & 15.55 & 15.11 & 0.44         \\
J163401.94$+$480940.2 & 2018 March 22, 26            & 16 34 01.94  &$+$48 09 40.20 & 19.54 & 19.21 & 0.33     \\
S1                    & 2018 March 22                & 16 34 04.24  &$+$48 11 32.47 & 19.65 & 18.77 & 0.88       \\ 
S2                    & 2018 March 22                & 16 33 50.78  &$+$48 10 09.78 & 19.04 & 17.86 & 1.18         \\
S3                    & 2021 April 09                & 16 33 31.81  &$+$48 04 31.89 & 19.73 & 18.34 & 1.39         \\
S4                    & 2021 April 09                & 16 34 01.24  &$+$48 08 36.92 & 18.25 & 17.30 & 0.95 \\

\hline
\multicolumn{7}{l}{The SDSS DR14 catalog~\citep{Abolfathi2018ApJS..235...42A} has been used for getting optical positions and apparent magnitudes of the sources  }\\
\multicolumn{7}{l}{and their comparison stars.}\\
\multicolumn{7}{l}{$^{*}$The USNO-A2.0 catalog~\citep{Monet1998AAS...19312003M} has been used in case of  non-availability of the SDSS `g-r' color. The `B-R' color has been}\\
\multicolumn{7}{l}{used in such cases.}\\
\end{longtable}
\end{center}
\twocolumn

\section{STATISTICAL ANALYSIS}
\label{sec4.0}

To guarantee the reliability of detection for microvariability events, multi-testing has mostly been used in recent years~\citep[e.g.,][]{Joshi2011MNRAS.412.2717J, Goyal2012A&A...544A..37G, Diego2014AJ....148...93D, Ojha2021MNRAS.501.4110O}. Therefore, for unambiguous detection of INOV in a DLC, we have used in the current work two different versions of $F$-test, which are the standard \emph{$F$-test} (hereafter $F^{\eta}$-test) and the power-enhanced \emph{$F$-test} (hereafter $F_{enh}$-test).
However, in the case of $F^{\eta}$-test, it is suggested that mismatching in the brightness levels among target AGN (in the current work, NLSy1 galaxy) and the two chosen steady comparison stars should be within $\sim$ 1-mag in order to avoid photon statistics and other random-noise terms~\citep[e.g., see][]{Howell1988AJ.....95..247H, Cellone2007MNRAS.374..357C, Goyal2012A&A...544A..37G}.
 Therefore, care was taken while selecting two (non-varying) comparison stars to be within 1-mag of the respective NLSy1s. Thus, for the present set of jetted-RLNLSy1s, the median magnitude mismatch of \emph{0.82} between the reference star (i.e., the comparison star with the closest match with the target AGN'S instrumental magnitude) and target NLSy1. The corresponding median values for the non-jetted-RLNLSy1s and the entire set of 23 NLSy1s are \emph{1.28} and \emph{0.91}, respectively. Additionally, while implementing the $F^{\eta}$-test, it is also crucial to use the correct RMS errors on the photometric data points due to underestimated magnitude errors by a factor ranging between 1.3 and 1.75, returned by the routines
in the data reduction software DAOPHOT and IRAF~\citep{Gopal-Krishna1995MNRAS.274..701G,
  Garcia1999MNRAS.309..803G, Sagar2004MNRAS.348..176S,
  Stalin2004JApA...25....1S, Bachev2005MNRAS.358..774B}. Therefore, the `$\eta$' value is taken here to be 1.54$\pm$0.05, { computed} using the data of 262 intra-night monitoring sessions of AGNs by~\citet{Goyal2013JApA...34..273G}.

Following~\citet{Goyal2012A&A...544A..37G}, $F^{\eta}$-statistics defined as

\begin{equation}
 \label{eq.fetest}
F_{s1}^{\eta} = \frac{\sigma^{2}_{(q-cs1)}} { \eta^2 \langle \sigma_{q-cs1}^2 \rangle}, 
\hspace{0.2cm} F_{s2}^{\eta} = \frac{\sigma^{2}_{(q-cs2)}} { \eta^2 \langle \sigma_{q-cs2}^2 \rangle}, 
\hspace{0.2cm} F_{s1-s2}^{\eta} = \frac{\sigma^{2}_{(cs1-cs2)}} { \eta^2 \langle \sigma_{cs1-cs2}^2 \rangle} 
\end{equation}

where $\sigma^{2}_{(q-cs1)}$, $\sigma^{2}_{(q-cs2)}$,  and $\sigma^{2}_{(cs1-cs2)}$ are the variances with $\langle \sigma_{q-cs1}^2 \rangle=\sum_ {i=1}^{N}\sigma^2_{i,~err}(q-cs1)/N$, $\langle \sigma_{q-cs2}^2 \rangle$, and $\langle \sigma_{cs1-cs2}^2 \rangle$ being the mean square (formal) rms errors of the individual data points in the `target NLSy1 - comparison star1', `target NLSy1 - comparison star2', and  `comparison star1 - comparison star2' DLCs, respectively.

The $F$-values were computed in this way for individual DLC using Eq.~\ref{eq.fetest}, and compared with the critical $F$-value, set, viz.,  $F^{(\alpha)}_{\nu}$, where $\alpha$ is the level of significance set by us for the $F^{\eta}$-test, and  $\nu$ ($= N_{j}-1$) is the degree of freedom for the DLC ($N_{j}$ being the number of data points in the DLC). The chance of a false INOV detection signal becomes lower for a smaller value of $\alpha$. Therefore, in the present work, similar to our previous work, two critical significance levels, $\alpha=$ 0.01 and $\alpha=$ 0.05 are set by us~\citep[e.g., see][]{Ojha2021MNRAS.501.4110O}. Following~\citet{Ojha2021MNRAS.501.4110O}, a NLSy1 is designated as variable for a given session according to this test if the computed value of $F^{\eta}$ is found to be greater than its $F_{c}(0.99)$. Table~\ref{NLSy1:tab_result} summarizes the computed $F^{\eta}$-values and the correspondingly inferred status of INOV detection for all the 53 sessions (columns 6 and 7).\par

The second flavor of $F$-test for INOV employed in the present study is the $F_{enh}$-test~\citep[e.g.,][]{Diego2014AJ....148...93D}. The statistical criteria for the $F_{enh}$-test can be described as

\begin{equation}
\label{Fenh_eq}  
\hspace{0.25cm} F_{{\rm enh}} = \frac{s_{{\rm NLSy1}}^2}{s_{\rm comb}^2}, \hspace{0.5cm} s_{\rm comb}^2=\frac{1}{(\sum _{j=1}^q N_j) - q}\sum _{j=1}^{q}\sum _{i=1}^{N_j}B_{j,i}^2 .
\end{equation}

here $s_{{\rm NLSy1}}^2$ is the variance of the `target NLSy1-reference star' DLC, while  $s_{\rm comb}^2$ is  the combined variance of `comparison star-reference star' DLC having $N_{j}$ data points and $q$ comparison stars, computed using  scaled square deviation $B_{{\rm j,i}}^2$ as

 \begin{equation}
\hspace{2.7cm} B_{j,i}^2=\omega _j(c_{j,i}-\bar{c}_{j})^2
 \end{equation}
 
 where, $c_{j,i}$'s is the `j$^{th}$ comparison star-reference star' differential instrumental magnitudes value and $\bar{c_{j}}$  represent the corresponding average value of the DLC for its $N_{j}$ data points. The scaling factor $\omega_ {j}$ is taken here as described in~\citet{Ojha2021MNRAS.501.4110O}.\par

 The principal feature of the $F_{enh}$-test is that it takes into account the brightness differences of the target AGN and the selected comparison stars, a frequently encountered problem with the $C$ and  $F$-statistics~\citep[e.g., see][]{Joshi2011MNRAS.412.2717J, Diego2014AJ....148...93D}. Thus $F_{enh}$-values were computed using Eq.~\ref{Fenh_eq} for individual DLCs and compared with the set critical values for this study (see above). Based upon this test, a NLSy1 DLC is assigned a designation ``variable (V)'' when the computed value of $F_{enh}$ found for `target NLSy1-reference star' DLC, is greater than its $F_{c}(0.99)$ (i.e., $F_{{\rm enh}} > F_{c}(0.99)$), and ``probable variable (PV)'' if same is found to be greater than $F_{c}(0.95)$ but less or equal to $F_{c}(0.99)$ (i.e., $F_{c}(0.95) < F_{\rm enh} \leq F_{c}(0.99)$). In Table~\ref{NLSy1:tab_result}, we tabulate the computed $F_{enh}$-values and the correspondingly inferred INOV status for our entire 53 sessions in columns 8 and 9.\par
 
\begin{table*}
  \begin{minipage}{195mm}
 \begin{center}   
 {\small
  \caption[caption]{ Details of the observations and the status of the INOV for the sample of 23 RLNLSy1 galaxies studied in this work (aperture radius used = 2$\times$FWHM).}

  \label{NLSy1:tab_result} \begin{tabular}{cccc ccccc cccr}
  \hline
  {RLNLSy1s} & Date(s)$^{a}$ &  T$^{b}$  & N$^{c}$  & Median$^{d}$ & {$F^{\eta}$-test} & {INOV}  & { {$F^{\eta}$-test}}& { Variability} & {$F_{enh}$-test} & {INOV} &{$\sqrt { \langle \sigma^2_{i,err} \rangle}$} & $\overline\psi^{g}_{s1, s2}$\\
  (SDSS name) & yyyy.mm.dd & (hrs) & & FWHM  & {$F_{s1}^{\eta}$},{$F_{s2}^{\eta}$} & status$^{e}$ & { {$F_{s1-s2}^{\eta}$}}& { status of}  & $F_{enh}$ & status$^{f}$  & (AGN-s)$^{g}$ & (\%) \\
  &&&& (arcsec) &           &{99\%}& { 99\%}& { s1$-$s2} &  &{99\%} &\\
  {(1)}&{(2)} & {(3)} & {(4)} & {(5)} & {(6)} & {(7)} & {(8)} & {(9)} & {(10)} & {(11)} &  {(12)}&  {(13)}\\
\hline
                                              \multicolumn{12}{c}{jetted NLSy1s}\\
\hline 
 J032441.20$+$341045.0 & (2016.11.22) & 4.42 & 56& 2.32& 14.34, 16.13 &  V,  V   &{ 00.82} &  { NV}  & 17.39 &   V  & 0.003 &  5.38\\  
                       &  2016.11.23  & 4.27 & 54& 2.13& { 03.77, 02.46} &  V,  V   &{ 00.70} &  { NV}  & { 05.42} &   V  & { 0.004} &  {2.79}\\
                       & (2016.12.02) & 4.41 & 44& 2.60& 85.80, 88.73 &  V,  V   &{ 00.87} &  NV  & 98.29 &   V  & 0.003 & 11.44\\
                       &  2017.01.03  & 3.00 & 39& 2.47& { 04.53, 08.53} &  V,  V   &{ 00.34} &  { NV}  & { 13.27} &   V  & { 0.004} &  {3.96}\\
                       &  2017.01.04  & 3.39 & 33& 2.45& { 21.60, 23.34} &  V,  V   &{ 00.44} &  { NV}  & { 49.35} &   V  &{  0.004} &  { 7.99}\\
 J081432.12$+$560958.7 & (2017.01.03) & 3.37 & 19& 2.69& 00.33, 00.34 &  NV, NV  &{ 00.16} &{  NV}  & 01.34 &   NV & 0.022 &    --\\
                       & (2017.11.20) & 4.52 & 32& 2.79& 00.86, 00.55 &  NV, NV  &{ 00.55} &{  NV}  & 05.28 &   NV & 0.031 &    --\\ 
 J084957.98$+$510829.0 & (2017.12.13) & 4.42 & 24& 2.83& 00.42, 00.51 &  NV, NV  &{ 01.06} &{  NV}  & 00.77 &   NV & 0.033 &    --\\
                       & (2019.04.08) & 3.04 & 13& 2.88& 00.66, 00.89 &  NV, NV  &{ 00.62} &{  NV}  & 00.62 &   NV & 0.032 &    --\\
 J090227.20$+$044309.0 & (2017.02.22) & 3.59 & 27& 2.38& 00.71, 00.57 &  NV, NV  &{ 00.25} &{  NV}  & 02.81 &   NV & 0.025 &    --\\
                       & (2017.12.14) & 5.65 & 39& 2.51& 00.33, 00.37 &  NV, NV  &{ 00.24} &{  NV}  & 01.38 &   NV & 0.024 &    --\\
 J094857.32$+$002225.6 & (2016.12.02) & 4.15 & 17& 2.58& 01.71, 01.88 &  NV, NV  &{ 00.16} &{  NV}  & 10.52 &   V  & 0.017 &  7.95\\
                       & (2017.12.21) & 5.19 & 33& 2.24& 13.95, 16.53 &  V , V   &{ 00.55} &{  NV}  & 25.26 &   V  & 0.012 & 16.42\\
 J095317.10$+$283601.5 & (2017.03.04) & 3.97 & 29& 2.41& 00.36, 00.35 &  NV, NV  &{ 00.48} &{  NV}  & 00.76 &   NV & 0.035 &    --\\
                       & (2020.11.21) & 3.25 & 11& 3.10& 00.77, 00.68 &  NV, NV  &{ 00.33} &{  NV}  & 02.31 &   NV & 0.035 &    --\\
 J104732.78$+$472532.0 & (2017.04.11) & 3.75 & 47& 0.98& 00.54, 00.53 &  NV, NV  &{ 00.53} &{  NV}  & 01.03 &   NV & 0.035 &    --\\
                       & (2018.03.12) & 3.82 & 15& 2.77& 00.60, 00.63 &  NV, NV  &{ 00.38} &{  NV}  & 01.58 &   NV & 0.028 &    --\\
 J122222.99$+$041315.9 &  2017.01.03  & 3.52 & 17& 2.38& 00.62, 00.30 &  NV, NV  &{ 00.91} &{  NV}  & 00.68 &   NV & 0.018 &    --\\  
                       &  2017.01.04  & 3.14 & 16& 2.36& 00.32, 00.37 &  NV, NV  &{ 00.16} &{  NV}  & 01.99 &   NV & 0.014 &    --\\ 
                       &  2017.02.21  & 4.44 & 41& 2.65& 00.74, 00.76 &  NV, NV  &{ 00.35} &{  NV}  & 02.13 &   V  & 0.020 &  6.36\\ 
                       & (2017.02.22) & 5.50 & 50& 2.59& 03.98, 03.60 &  V , V   &{ 00.61} &{  NV}  & 06.51 &   V  & 0.017 & 13.33\\ 
                       & (2017.03.04) & 4.93 & 39& 2.61& 00.72, 00.86 &  NV, NV  &{ 00.53} &{  NV}  & 01.36 &   NV & 0.019 &    --\\  
                       &  2017.03.24  & 3.94 & 39& 2.37& 00.93, 00.75 &  NV, NV  &{ 00.56} &{  NV}  & 01.66 &   NV & 0.020 &    --\\
 J130522.75$+$511640.2 & (2017.04.04) & 3.79 & 23& 2.57& 00.70, 00.66 &  NV, NV  &{ 00.24} &{  NV}  & 02.94 &   PV & 0.012 &    --\\
                       & (2019.04.25) & 3.11 & 22& 2.77& 01.56, 01.42 &  NV, NV  &{ 00.39} &{  NV}  & 04.12 &   PV & 0.018 &    --\\
 J142114.05$+$282452.8 & (2018.05.10) & 4.06 & 26& 2.83& 00.55, 00.56 &  NV, NV  &{ 00.35} &{  NV}  & 01.60 &   NV & 0.019 &    --\\
                       & (2019.05.27) & 3.31 & 18& 2.68& 00.86, 00.86 &  NV, NV  &{ 00.36} &{  NV}  & 02.47 &   PV & 0.021 &    --\\
 J144318.56$+$472556.7 & (2018.03.11) & 3.05 & 19& 3.15& 00.54, 00.56 &  NV, NV  &{ 00.21} &{  NV}  & 02.59 &   PV & 0.022 &    --\\
                       & (2018.03.23) & 3.13 & 23& 2.33& 00.35, 00.36 &  NV, NV  &{ 00.35} &{  NV}  & 01.00 &   NV & 0.018 &    --\\
 J150506.48$+$032630.8 & (2017.03.25) & 5.21 & 41& 2.08& 00.60, 00.59 &  NV, NV  &{ 00.58} &{  NV}  & 01.04 &   NV & 0.028 &    --\\
                       & (2018.04.12) & 3.05 & 19& 2.55& 00.67, 00.63 &  NV, NV  &{ 00.80} &{  NV}  & 00.84 &   NV & 0.032 &    --\\
 J154817.92$+$351128.0 & (2018.05.17) & 3.00 & 19& 3.08& 00.40, 00.38 &  NV, NV  &{ 00.38} &{  NV}  & 01.05 &   NV & 0.008 &    --\\
                       & (2019.05.08) & 3.24 & 14& 2.79& 00.60, 00.65 &  NV, NV  &{ 00.30} &{  NV}  & 01.98 &   NV & 0.017 &    --\\
 J164442.53$+$261913.3 & (2017.04.03) & 4.37 & 37& 2.50& 01.44, 01.28 &  NV, NV  &{ 00.41} &{  NV}  & 03.53 &   V  & 0.011 &  5.41\\
                       & (2019.04.26) & 3.22 & 24& 2.27& 03.06, 03.74 &  V , V   &{ 00.48} &{  NV}  & 06.40 &   V  & 0.011 &  7.50\\
 J170330.38$+$454047.3 & (2017.06.03) & 3.76 & 37& 2.41& 00.75, 00.67 &  NV, NV  &{ 00.64} &{  NV}  & 01.17 &   NV & 0.004 &    --\\
                       & (2019.03.25) & 3.13 & 45& 4.45& 00.63, 01.08 &  NV, NV  &{ 00.90} &{  NV}  & 00.70 &   NV & 0.010 &    --\\
\hline										  	  
                         \multicolumn{12}{c}{non-jetted NLSy1s}\\		  	  
\hline										  	  
 J085001.17$+$462600.5 & (2017.01.04) & 3.26 & 13& 2.28& 00.39, 00.50 &  NV, NV  &{ 00.50} &{  NV}  & 00.78 &   NV & 0.026 &    --\\
                       & (2017.12.15) & 3.69 & 20& 2.89& 00.80, 00.74 &  NV, NV  &{ 00.36} &{  NV}  & 02.22 &   NV & 0.032 &    --\\
 J103727.45$+$003635.6 & (2018.03.11) & 3.30 & 11& 3.12& 01.18, 01.22 &  NV, NV  &{ 01.08} &{  NV}  & 01.09 &   NV & 0.030 &    --\\
                       & (2021.04.08) & 4.96 & 13& 2.42& 01.54, 01.77 &  NV, NV  &{ 01.00} &{  NV}  & 01.53 &   NV & 0.043 &    --\\
 J111005.03$+$365336.2 & (2018.03.23) & 3.22 & 44& 0.90& 00.34, 00.35 &  NV, NV  &{ 00.47} &{  NV}  & 00.72 &   NV & 0.027 &    --\\
                       & (2019.01.13) & 3.13 & 11& 3.07& 02.61, 02.54 &  NV, NV  &{ 01.18} &{  NV}  & 02.11 &   NV & 0.042 &    --\\
 J113824.53$+$365327.2 & (2017.04.17) & 4.32 & 20& 2.11& 00.32, 00.40 &  NV, NV  &{ 00.21} &{  NV}  & 01.58 &   NV & 0.028 &    --\\
                       & (2018.03.23) & 4.31 & 21& 2.53& 00.36, 00.38 &  NV, NV  &{ 00.16} &{  NV}  & 02.26 &   NV & 0.029 &    --\\
 J120014.08$-$004638.7 & (2018.03.12) & 3.83 & 28& 2.72& 00.31, 00.22 &  NV, NV  &{ 00.19} &{  NV}  & 01.63 &   NV & 0.011 &    --\\
                       & (2018.05.11) & 3.13 & 16& 2.97& 00.13, 00.25 &  NV, NV  &{ 00.33} &{  NV}  & 00.39 &   NV & 0.018 &    --\\
 J124634.65$+$023809.1 & (2017.04.03) & 3.77 & 18& 2.50& 00.21, 00.36 &  NV, NV  &{ 00.46} &{  NV}  & 00.46 &   NV & 0.024 &    --\\
                       & (2018.04.12) & 3.72 & 21& 2.71& 00.59, 00.66 &  NV, NV  &{ 00.31} &{  NV}  & 01.89 &   NV & 0.018 &    --\\
 J163323.59$+$471859.0 & (2017.05.20) & 4.33 & 36& 2.26& 00.85, 00.76 &  NV, NV  &{ 00.23} &{  NV}  & 03.76 &   V  & 0.007 &    2.56\\
                       & (2019.03.20) & 3.69 & 33& 2.66& 01.61, 01.79 &  NV, NV  &{ 00.36} &{  NV}  & 04.43 &   V  & 0.016 &    9.52\\  
 J163401.94$+$480940.2 & (2018.03.23) & 3.04 & 34& 0.98& 00.62, 00.71 &  NV, NV  &{ 00.39} &{  NV}  & 01.61 &   NV & 0.012 &    --\\
                       & (2021.04.09) & 4.58 & 12& 2.12& 01.27, 01.63 &  NV, NV  &{ 00.34} &{  NV}  & 03.76 &   PV & 0.043 &    --\\

 \hline
  \multicolumn{12}{l}{$^{a}$Date(s) of the monitoring session(s). The dates given inside parentheses refer to the sessions we have used here for estimating the INOV duty cycle}\\
  \multicolumn{12}{l}{(e.g., see text in Sect.~\ref{sec 4.1}). $^{b}$Duration of the monitoring session in the observed frame. $^{c}$Number of data points in the DLCs of the monitoring session. }\\
  \multicolumn{12}{l}{$^{d}$Median seeing (FWHM in arcsec) for the session. $^{e,~f}$INOV status inferred from F$^{\eta}$ and F$_{enh}$ tests, with V = variable , i.e. confidence level $\geq$ 99\%;}\\
  \multicolumn{12}{l}{PV = probable variable, i.e. $95-99$\% confidence level; NV = non-variable, i.e. confidence level $<$ 95\%.}\\
\multicolumn{12}{l}{$^{g}$Mean amplitude of variability in the two DLCs of the target NLSy1 (i.e., relative to the two chosen comparison stars).}\\
    \end{tabular}  
 }              
\end{center}
 \end{minipage} 
    \end{table*}

\begin{table*}

 {\small
   \caption{The DC and $\overline{\psi}$ of INOV, for the sample of 23 RLNLSy1 galaxies studied in this work, based on the $F_{enh}$-test and  $F^{\eta}$-test.}
   
 \label{NLSy1:DCy_result}
\centering
 \begin{tabular}{cccccccc}
   \hline
    RLNLSy1s & No. of Sources & \multicolumn{2}{ c }{$F_{enh}$-test} & \multicolumn{2}{ c }{$F^{\eta}$-test} & \multicolumn{2}{ c }{Median black hole mass}\\
     \hline
                            &     &{$^{\star}$DC}   &{$^{\star}\overline{\psi}^{\dag}$} &   {$^{\star}$DC}   &  {$^{\star}\overline{\psi}^{\dag}$} & log ($M_{BH}/M_{\sun}$)
                            \\
                             &    &({\%})         &  ({\%})                  &  ({\%})  &     ({\%}) &         \\
  \hline
  {jetted-RLNLSy1s}    & 15         &  18 (30)$^{\bot}$  &   09 (07)$^{\bot}$    &   12 (30)$^{\bot}$    &   11 (05)$^{\bot}$ & 7.72\\ 
  {non-jetted-RLNLSy1s} & 8        &  05 (16)$^{\bot}$  &   09 (01)$^{\bot}$    &   00 (16)$^{\bot}$    &   --  & 7.42\\ \hline
  {J-$\gamma$-RLNLSy1s}& 8        &  34 (16)$^{\bot}$  &   10 (07)$^{\bot}$    &   29 (16)$^{\bot}$    &   11 (05)$^{\bot}$ & 7.59\\ 
  {J-RLNLSy1s} & 7                 &  00 (14)$^{\bot}$  &   --   &   00 (14)$^{\bot}$    &   --  & 7.73 &\\ 
       
  \hline
  
  \multicolumn{7}{l}{$^{\star}$We used the 46 sessions for this estimation, as explained in Sect.~\ref{sec 4.1}. $^{\dag}$The mean value for all the DLCs belonging to the type `V'.}\\
  \multicolumn{7}{l}{$^{\bot}$Values inside parentheses are the number of observing sessions used to estimate the parameters DC or $\overline{\psi}$.}\\

 \end{tabular}  
 }             

\end{table*}

\subsection{Computation of INOV duty cycle and amplitude of variability}
\label{sec 4.1}

To compute the duty cycle (DC) of INOV for the present sets of RLNLSy1s, we have adopted, following the definition given by~\citet{Romero1999A&AS..135..477R}~\citep[see, also][]{Stalin2004JApA...25....1S}

\begin{equation} 
\hspace{2.5cm} DC = 100\frac{\sum_{j=1}^n R_j(1/\Delta t_j)}{\sum_{j=1}^n (1/\Delta t_j)} 
\hspace{0.1cm}{\rm per~cent} 
\label{eqno1} 
\end{equation} 

where $\Delta t_j = \Delta t_{j,~observed}(1+$z$)^{-1}$ ($z$ being the redshift of the target NLSy1 galaxy in current study) is the target AGN's redshift corrected time duration of the $j^{th}$ monitoring session~\citep[see details in][]{Ojha2020MNRAS.493.3642O}. For $j^{th}$ session, $R_j$ is considered to be 1 in Eq.~\ref{eqno1} only when INOV is detected, otherwise taken to be zero. Note that to avoid introducing bias, we have used only 2 sessions for each AGN. For sources observed in more than 2 sessions (e.g., see Table~\ref{NLSy1:tab_result}), the computation of DC used only the longest two sessions, as pointed out in~\citet{Ojha2020MNRAS.493.3642O, Ojha2021MNRAS.501.4110O}. The computed INOV duty cycles (DCs) for the different sets of RLNLSy1s are listed in Table~\ref{NLSy1:DCy_result}, based on two statistical tests .\par

To compute the peak-to-peak amplitude of INOV ($\psi$) detected in a given DLC,
we followed the definition given by~\citet{Heidt1996A&A...305...42H}

\begin{equation} 
\hspace{2.5cm} \psi= \sqrt{({H_{max}}-{H_{min}})^2-2\sigma^2}
\end{equation} 

with $H_{min,~max}$ = minimum (maximum) values in the DLC of target NLSy1 relative to steady comparison stars and $\sigma^2 = \eta^2\langle\sigma^2_{NLSy1-s}\rangle$, where, $\langle\sigma^2_{NLSy1-s}\rangle$ is the mean square (formal) rms errors of individual data points. The mean value of ($\overline{\psi}$) for different sets (e.g., see Table~\ref{NLSy1:DCy_result}) of RLNLSy1 galaxies is computed by taking average of the computed $\psi$ values for the DLCs belonging to the ``V'' category. In Table~\ref{NLSy1:DCy_result}, we have summarized the computed $\overline{\psi}$ values based on the two statistical tests for the different sets of RLNLSy1s in our sample.

\section{Results and discussion}
\label{sec 5.0}

The INOV characterization of RLNLSy1 galaxies presented here is likely to be more representative in comparison to the previous studies based on significantly smaller samples ~\citep{Liu2010ApJ...715L.113L, Paliya2013MNRAS.428.2450P, Kshama2017MNRAS.466.2679K}. Also, we have paid particular attention to guarding against the possibility of spurious INOV claims arising from a varying flux contribution to the aperture photometry from the host galaxy of the AGN caused due to seeing disc variation during the session. As pointed out by~\citet{Cellone2000AJ....119.1534C}, under such circumstances, false claims of INOV can result in $low-z$ AGNs. Based on recent deep imaging studies of NLSy1 galaxies by~\citet{2020MNRAS.492.1450O}, it can be inferred that any variable contamination arising from the host galaxy is very unlikely to matter when studying the variability of AGNs at least at z $>$ 0.5. From Table~\ref{NLSy1:tab_result}, it is seen that the INOV detection ($\psi > 3$\% ) has occurred for just 3 sources in our sample having z $<$ 0.5. These are: J032441.20$+$341045.0 (at $z = 0.06$; 4 sessions), J164442.53$+$261913.3 (at $z = 0.14$; two sessions), and J163323.59$+$471859.0 ($z = 0.12$, one session). However, since the seeing disk remained steady in all these sessions (Fig.~\ref{fig:lurve 1}), except for the flickering in the PSF for the 3-4 points only in the case of J164442.53$+$261913.3 for its first session (Fig.~\ref{fig:lurve 4}) and a non-negligible systematic PSF variation in the case of J163323.59$+$471859.0 (Fig.~\ref{fig:lurve 6}). A closer checkup of these two intranight sessions shows that either PSF remained fairly steady during the time of AGN's flux variations (Fig.~\ref{fig:lurve 4}) or the gradients in the DLCs of the target AGN are seen to be anticorrelated with systematic variations in the PSF (Fig.~\ref{fig:lurve 6}). This is opposite to what is
expected in case the aperture photometric measurements were significantly contaminated by the underlying galaxy~\citep[see][]{Cellone2000AJ....119.1534C}. Therefore, the possibility of a significant variation in the fractional contribution from the host galaxy can be safely discounted. We thus conclude that the present cases of INOV detection for $low-z$ RLNLSy1s are genuine and not artifacts of seeing disk variation through the monitoring sessions. \par 
From Table~\ref{NLSy1:DCy_result}, it is seen that the $F_{enh}$-test resulted in the DCs of 18\% and 5\% for the jetted and non-jetted-RLNLSy1s sets; however, the DCs of 12\% and 0\% are estimated based on the conservative $F^{\eta}$-test for the same sets. Thus, regardless of which of the two statistical tests is applied, the INOV DCs for the jetted sample is higher than the non-jetted sample. Since we are using only two steady comparison stars in this study, therefore power-enhanced F-test becomes similar to that of the scaled
F-test. However, it has been suggested in~\citet{Goyal2013MNRAS.435.1300G} that in such a condition power-enhanced F-test does not follow the standard F-distribution.
Therefore, for further discussion and conclusion, we will be using our results based upon $F^{\eta}$-test. Additionally, in the case of powerful quasars/blazars, such DCs characterize non-blazars, including weakly polarised flat-radio-spectrum (i.e., radio-beamed) quasars~\citep{Goyal2013MNRAS.435.1300G, Gopal-Krishna2018BSRSL..87..281G}. It is further seen from Table~\ref{NLSy1:DCy_result} that a higher DC~\citep[ $\sim$30\%, i.e., approaching blazar-like values, e.g., see][]{Goyal2013MNRAS.435.1300G} is only exhibited by the $\gamma$-ray detected subset, which consists of 8 RLNLSy1s, all of which belong to the jetted category. This independently reinforces the premise that $\gamma$-ray detected NLSy1 galaxies emit blazar-like compact radio jets, which are likely conspicuous due to relativistic beaming. The absence of $\gamma$-ray detection among the non-jetted RLNLSy1s is interesting because it suggests the non-detection of jets in these sources. It is probably related to relativistic dimming (due to misalignment) rather than due to the jets being beamed but still too small physically to be resolved by VLBA. This inference is based on the currently popular notion that $\gamma$-rays in AGN primarily arise from the vicinity of the central engine, i.e., the base of the jets~\citep[e.g., see][]{Neronov2015NatPh..11..664N}. It would then appear that the radio jets in the non-jetted RLNLSy1s either are not strongly beamed or weak. In this context, we also note from Table~\ref{NLSy1:DCy_result}, that the DC of 8 J-$\gamma$-RLNLSy1s is 29\% while none of the sources in the sample of 7 J-RLNLSy1s has shown INOV. As also noted in Sect.~\ref{section_2.0} that the VLBA detected jets might not be relativistically beamed, which may result in the non-detection of INOV in these sources. The higher INOV DC in 8 J-$\gamma$-RLNLSy1s may be related to the relativistic beaming of jets in these sources, and the mere presence of jets does not guarantee an INOV detection in NLSy1 galaxies. For instance, J032441.20$+$341045.0 has lowest polarization value (e.g., see Table~\ref{tab:source_info}) and radio loudness value~\citep[e.g., see][]{Zhou2007ApJ...658L..13Z} among 8 J-$\gamma$-RLNLSy1s but shows strong INOV activity which perhaps could be due to very high jets speed (e.g., see Table~\ref{tab:source_info}). This supports our above argument that relativistic beaming plays a significant role for INOV in the case of gamma-ray detected radio-loud NLSy1s.\par

From Table~\ref{NLSy1:DCy_result}, the DC estimates, based upon $F^{\eta}$-test for the samples of jetted and non-jetted RLNLSy1s, have contrasting differences. Thus, our results suggest that the jetted-RLNLSy1s sample shows higher DC in comparison to the non-jetted-RLNLSy1s. However, as noted above that, instead of the mere presence of a jet, relativistic beaming seems to play a dominant role for INOV in the case of low-luminous high accreting AGNs such as NLSy1 galaxies. As has been emphasized in Sect.~\ref{sec1.0} that the central engine of NLSy1s operates in a regime of a higher accretion rate, and contributions from their host galaxies are prominent in the case of lower redshift sources (see Sect.~\ref{sec3.2}). However, our method of analysis for INOV detections is very unlikely to be affected by the host galaxy's contributions (hence spurious INOV detection) from the present sources, but due to higher accretion rates of NLSy1s, a relative enhancement in the AGN's optical emission (i.e., thermal component) as compared to its synchrotron emission is expected~\citep[e.g., see][]{Zhou2007ApJ...658L..13Z, Paliya2014ApJ...789..143P}.
Since AGN's optical emission is likely to be less amenable to being variable in comparison to synchrotron emission, resulting from the Doppler boosted synchrotron jet; therefore the amplitude of INOV is expected to be suppressed by this thermal contamination. Thus, the DC of the jetted and non-jetted RLNLSy1s samples would be more robust once it becomes possible to subtract out thermal contamination originating in NLSy1s from the disc due to their higher Eddington accretion rates. Furthermore, a relatively lower DC of the sample of jetted-RLNLSy1s might be either due to the sub-luminal speed of jets~\citep[e.g., see][]{Ojha2019MNRAS.483.3036O} or due to their primarily misaligned relativistic jets towards the observer line of site~\citep{Berton2018A&A...614A..87B}. Therefore, we categorized our sample of jetted-RLNLSy1s into the subsamples of J-$\gamma$-RLNLSy1s (8 sources) and J-RLNLSy1s (7 sources) based on their detections in $\gamma$-ray by  {\it Fermi}-LAT, where detection of $\gamma$-ray emission supports the scenario of presence of Doppler boosted relativistic jets~\citep{Abdo2009ApJ...699..976A, Abdo2009ApJ...707..727A, Abdo2009ApJ...707L.142A, Foschini2010ASPC..427..243F, Foschini2011nlsg.confE..24F, D'Ammando2012MNRAS.426..317D, D'Ammando2015MNRAS.452..520D, Yao2015MNRAS.454L..16Y, Paliya2018ApJ...853L...2P, Yang2018MNRAS.477.5127Y, Yao2019MNRAS.487L..40Y}. It can be seen in Table~\ref{NLSy1:DCy_result} that we found a lack of INOV detection based upon the $F^{\eta}$-test in the J-RLNLSy1s subsample in contrast to the DC of $\sim$ 29\% for J-$\gamma$-RLNLSy1s subsample, consistent with the result of~\citet{Ojha2019MNRAS.483.3036O}, where based upon a small sample of three NLSy1s, it was found out that superluminal motion in the radio jet could be a robust diagnostic of INOV.\par
To further confirm the above scenario, we compiled the apparent jet speed of our J-$\gamma$-RLNLSy1s from the literature. We found that out of 8 members of J-$\gamma$-RLNLSy1s, six members have available apparent jet speeds. These six J-$\gamma$-RLNLSy1s are: J032441.20$+$341045.0, J084957.98$+$510829.0, J094857.32$+$002225.6, J150506.48$+$032630.8, with $v_{app}/c$ of $9.1\pm0.3$, $6.6\pm0.6$, $9.7\pm1.1$, $0.1\pm0.2$, respectively~\citep[e.g., see][]{Lister2019ApJ...874...43L}, and J122222.99$+$041315.9, J164442.53$+$261913.3, with $v_{app}/c$ of $0.9\pm0.3$, $>$1.0, respectively~(\citealp[e.g., see][]{Lister2016AJ....152...12L},~\citealp[and][]{Doi2012ApJ...760...41D}). Thus with the current INOV study and available jet speed of J-$\gamma$-RLNLSy1s, a correlated INOV detection with the superluminal motion in the radio jet is inferred, except for J084957.98$+$510829.0. Although, unfortunately, we could not find any INOV in both $>$ 3 hrs long monitoring sessions of the source, J084957.98$+$510829.0 in the present study, this source has shown in the past a fading by $\sim$ 0.2 mags in its INOV study within just $\sim$ 15 minutes during its high $\gamma$-ray active phase~\citep[see figure 6 of][]{Maune2014ApJ...794...93M}. Other than this instance, this source previously had also shown significant INOV in all its six intra-night sessions~\citep[e.g., see][]{Paliya2016ApJ...819..121P}. Therefore, the non-detection of INOV in our current study may be due to currently undergoing the quiescent $\gamma$-ray phase of this source. Nonetheless, the above correlation could be firmly established once a more comprehensive INOV database and apparent jet speeds of J-$\gamma$-RLNLSy1s become available.\par

On the other hand, in the case of RLNLSy1s (with $R > 100$), where most of the electromagnetic emission (radio, optical, X-ray, and $\gamma$-ray) supposedly comes from their jets, it is suggested that more massive black holes are more amenable to launch powerful relativistic jets~\citep{Urry2000ApJ...532..816U, 2007A&A...476..723H, Chiaberge2011MNRAS.416..917C, 2016MNRAS.460.3202O}. Therefore, we compared the median black hole masses of jetted-RLNLSy1s and non-jetted-RLNLSy1s samples, derived based upon single-epoch optical spectroscopy virial method (e.g., see second last column of Table~\ref{tab:source_info}). This has resulted in a nominal difference with the median values of log ($M_{BH}$/$M_{\sun}$) of \emph{7.72} and \emph{7.42} for the jetted-RLNLSy1s and non-jetted-RLNLSy1s samples (see last column of Table~\ref{NLSy1:DCy_result}), respectively. A contrasting difference has not been found between the black hole masses of the two sets, which may be either due to smaller sample sizes or due to the use of the single-epoch optical spectroscopy virial method, which is suggested to have a systematic underestimation while estimating black hole masses~\citep{Decarli2008MNRAS.386L..15D, Marconi2008ApJ...678..693M, Calderone2013MNRAS.431..210C, Viswanath2019ApJ...881L..24V, Ojha2020ApJ...896...95O}. Therefore, to firmly establish the above scenario, estimation of black hole masses of a large sample of jetted and non-jetted RLNLSy1s is needed with the method that is less likely to be affected by underestimation of black hole masses such as the standard Shakura-Sunyaev accretion-disc model method~\citep[][]{Calderone2013MNRAS.431..210C, Viswanath2019ApJ...881L..24V}.\par

\section{Conclusions}
\label{sec6.0}
To quantify the role of the absence/presence of radio jets for INOV in the case of RLNLSy1s, we have carried out a systematic INOV study based on an unbiased sample of 23 RLNLSy1s. Among them, 15 RLNLSy1s have confirmed detection of jets (jetted), and the remaining 8 RLNLSy1s have no detection of jets (non-jetted) with the VLBA observations. Our study spans 53 sessions of a minimum 3-hour duration each. The main conclusions from this work are as follows:\\

\begin{enumerate}
    \item We estimated the INOV DC based upon $F^{\eta}$-test for the sample of jetted RLNLSy1s to be 12\%, however, none of the sources showed INOV in the sample of non-jetted RLNLSy1s, at the 99\% confidence level for a typical threshold $\psi > 3$\%.

    \item Among the jetted RLNLSy1s, the DC for jetted $\gamma$-ray detected RLNLSy1s  is found to be  29\% in contrast to null INOV detection in the case of non-$\gamma$-ray detected RLNLSy1s. It suggests that the INOV detection in RLNLSy1 galaxies does not solely depend on the presence of radio jets, but relativistic beaming plays a dominant role.  
    
    \item The predominance of beamed jet for INOV is also supported in our study based on the correlation of the INOV detection with the apparent jet-speed available for 6 jetted $\gamma$-ray detected RLNLSy1s.

    \item  The higher DC of $\sim$ 30\%, approaching blazar-like DC, is only exhibited by the $\gamma$-ray detected subset, suggests that $\gamma$-ray detected NLSy1 galaxies emit blazar-like compact radio jets in which relativistic jet motion (speed) and/or small jet's angles to the observer's line of sight seems to be correlated with the presence of INOV.
\end{enumerate}

For further improvement, it will be helpful to enlarge the sample and conduct similar systematic INOV studies for other subclasses of AGN with and without a confirmed jet, along with the proper estimate of apparent jet speeds.

\section*{Acknowledgements}
{ We are thankful to the anonymous referee for providing comments and suggestions, which helped us improve the manuscript considerably.}
This research is part of the DST-SERB project under grant no. EMR/2016/001723. VKJ and HC acknowledge the financial support provided by DST-SERB for this work. We are very grateful to Prof. Gopal-Krishna for his helpful scientific discussions and important suggestions for this work. We thank the concerned ARIES and IIA staff for assistance during the observations obtained at the 3.6-m Devasthal Optical Telescope (DOT) and the 2.01-m Himalayan Chandra Telescope (HCT), which are the national facilities run and managed by ARIES and IIA, respectively, as autonomous Institutes under Department of Science and Technology, Government of India.

\section*{Data availability}

The data from the \href{https://www.aries.res.in/facilities/astronomical-telescopes}{ARIES telescopes} and the 2.01-m  \href{https://www.iiap.res.in/iao/cycle.html}{HCT telescope of IIA} used in this paper will be shared on reasonable request to the corresponding author.

\begin{figure*}
\centering
\includegraphics[height=21cm, width=19cm]{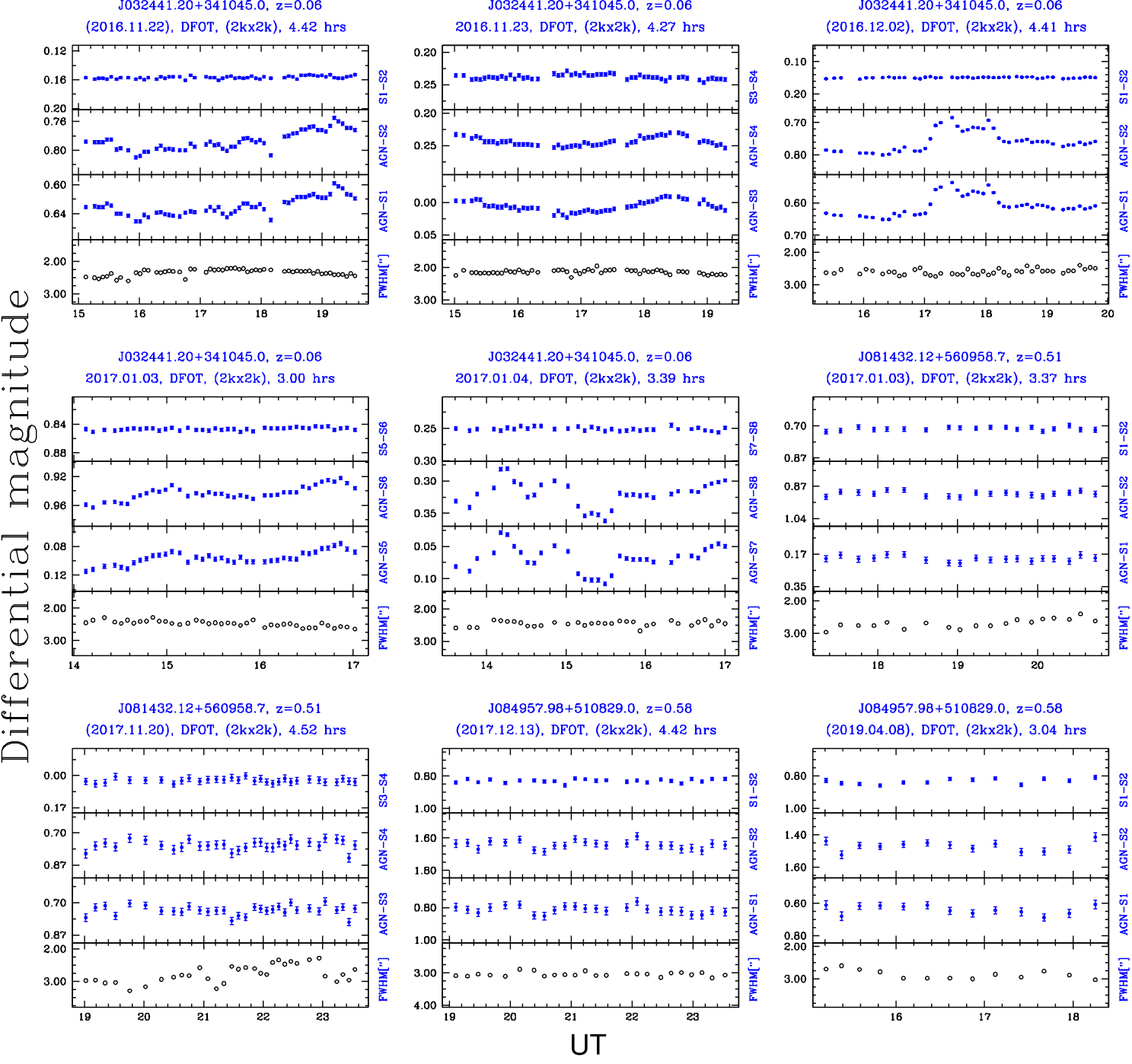}
\caption[]{ The differential light curves (DLC) for the first 3 jetted-RLNLSy1s from our sample of 15 jetted-RLNLSy1s are shown here. The name of the RLNLSy1 galaxy, the redshift (z), the name of the telescope used, and the duration of the observations are shown on the top of each panel. The light curves generated from the data obtained on the dates given inside parentheses at the top of each panel were used for the statistical analysis. In each panel, the DLC on the top is the instrumental magnitude difference between two non-variable comparison stars, the two DLC in the middle are prepared using the NLSy1 and the two comparison stars, respectively, while the bottom DLC shows the variation of the seeing conditions (FWHM in arcseconds)  during the monitoring session.}

\label{fig:lurve 1}
\end{figure*}

\begin{figure*}
\ContinuedFloat

\includegraphics[height=21cm, width=19cm]{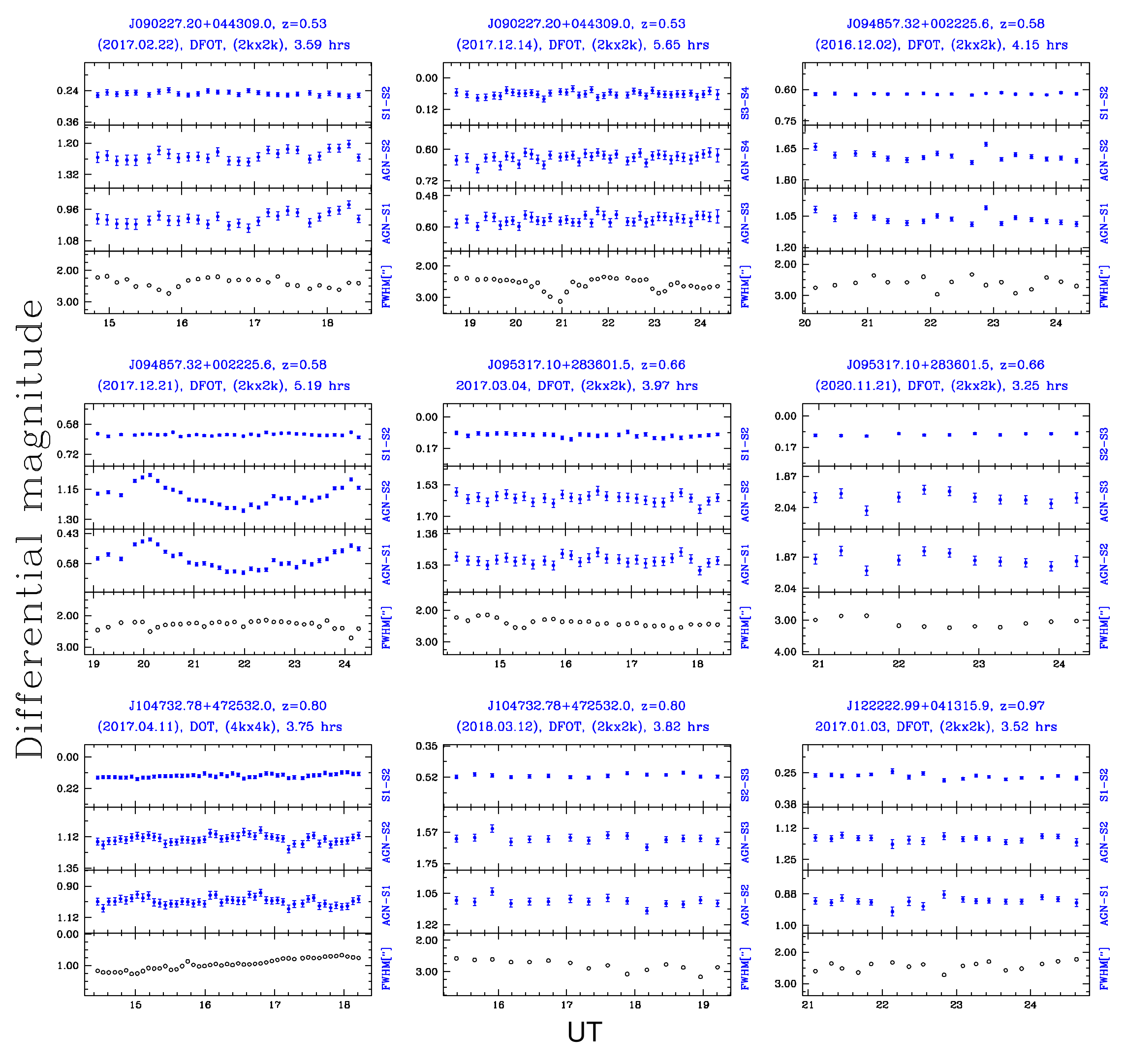}
\caption[]{(Continued) DLC for the subsequent 5 jetted-RLNLSy1s from the current sample of 15 jetted-RLNLSy1s.}
\label{fig:lurve 2}
\end{figure*}

\begin{figure*}
\ContinuedFloat

\centering
\includegraphics[height=21cm, width=19cm]{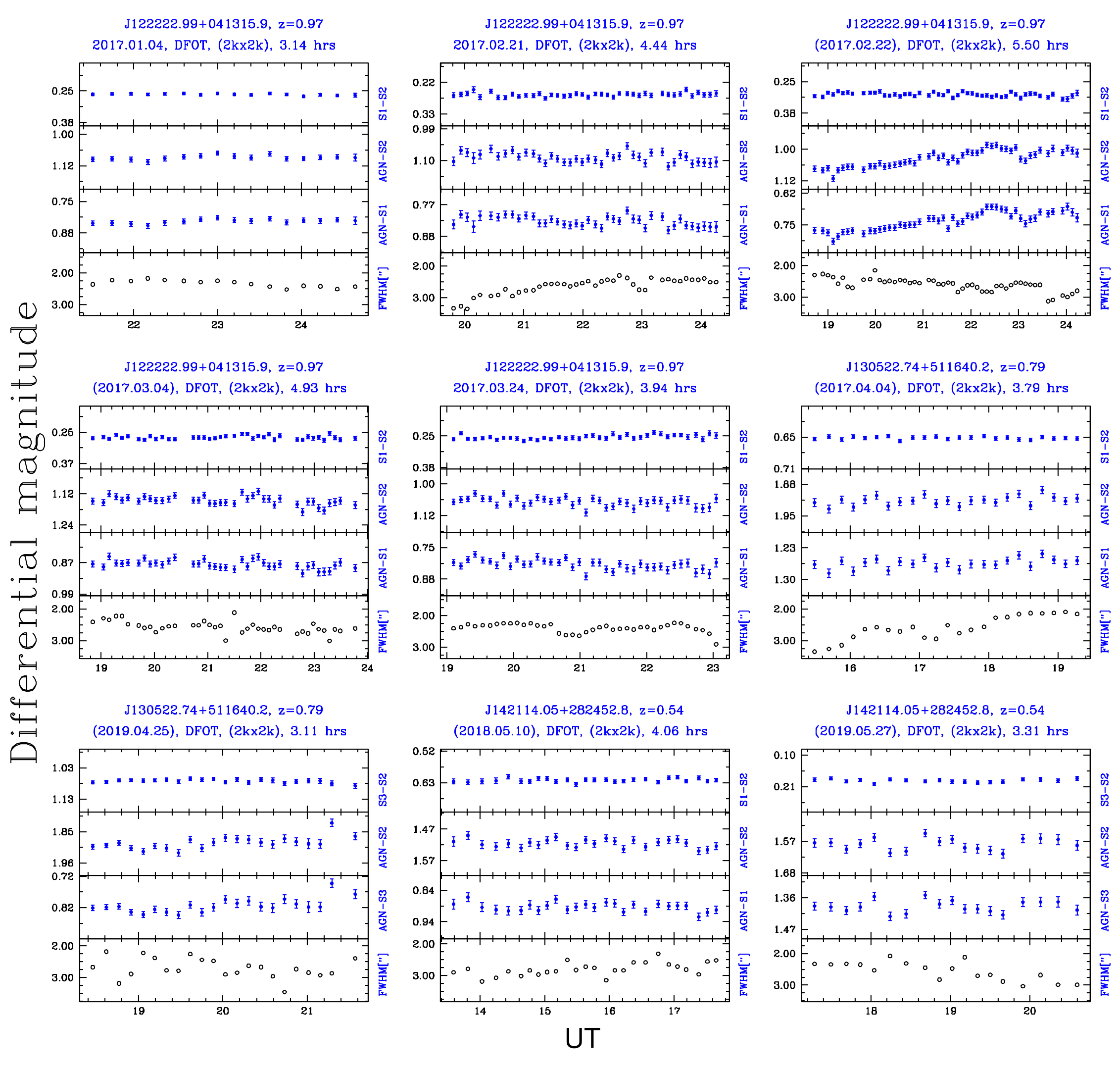}
\caption[]{(Continued) DLC for another 2 jetted-RLNLSy1s from the current sample of 15 jetted-RLNLSy1s.}
\label{fig:lurve 3}
\end{figure*}

\begin{figure*}
\ContinuedFloat

\includegraphics[height=21cm, width=19cm]{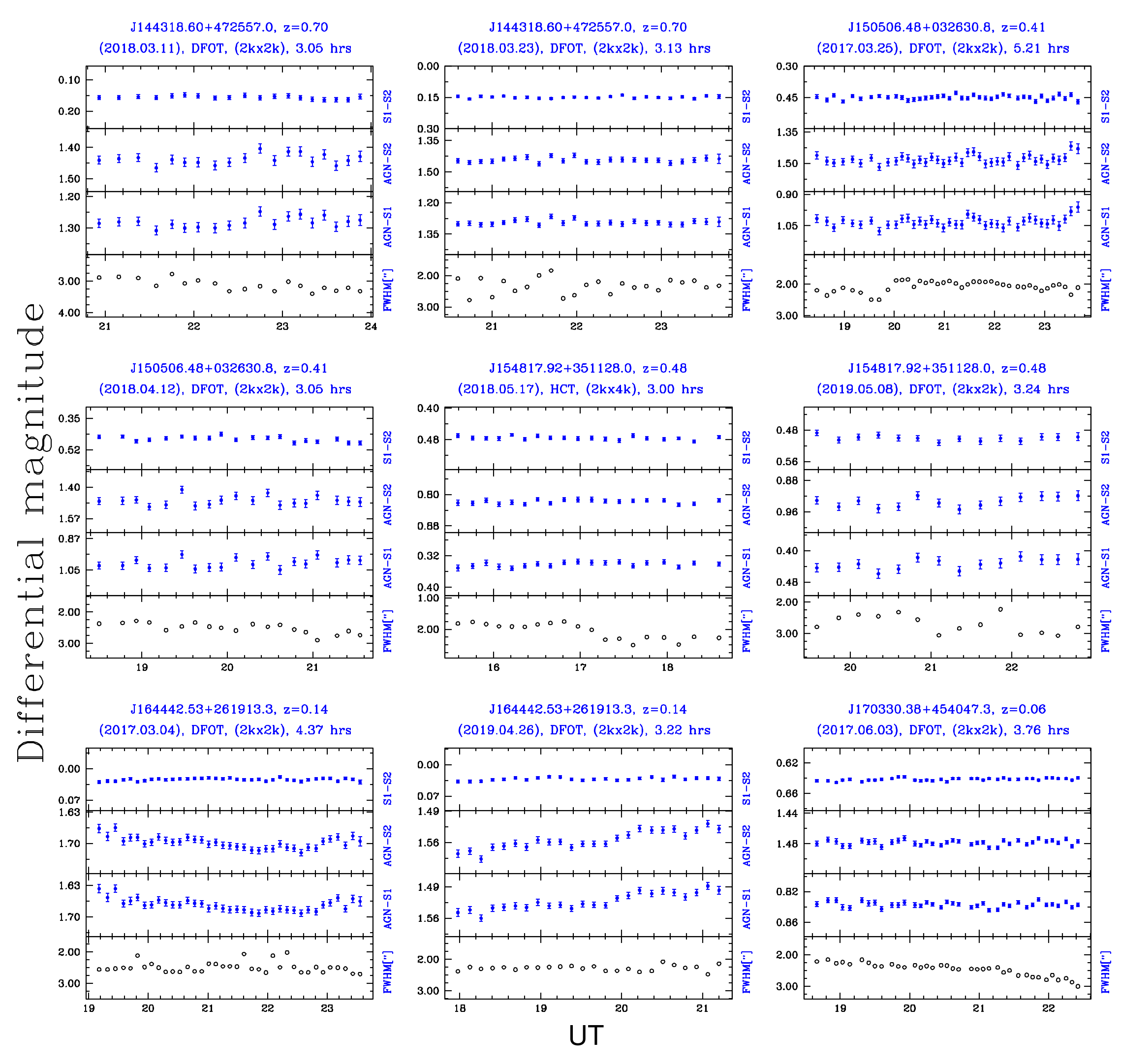}

\caption[]{ (Continued) DLC for the final 5 jetted-RLNLSy1s from the current sample of 15 jetted-RLNLSy1s.}
\label{fig:lurve 4}
\end{figure*}

\begin{figure*}
\includegraphics[height=21cm, width=19cm]{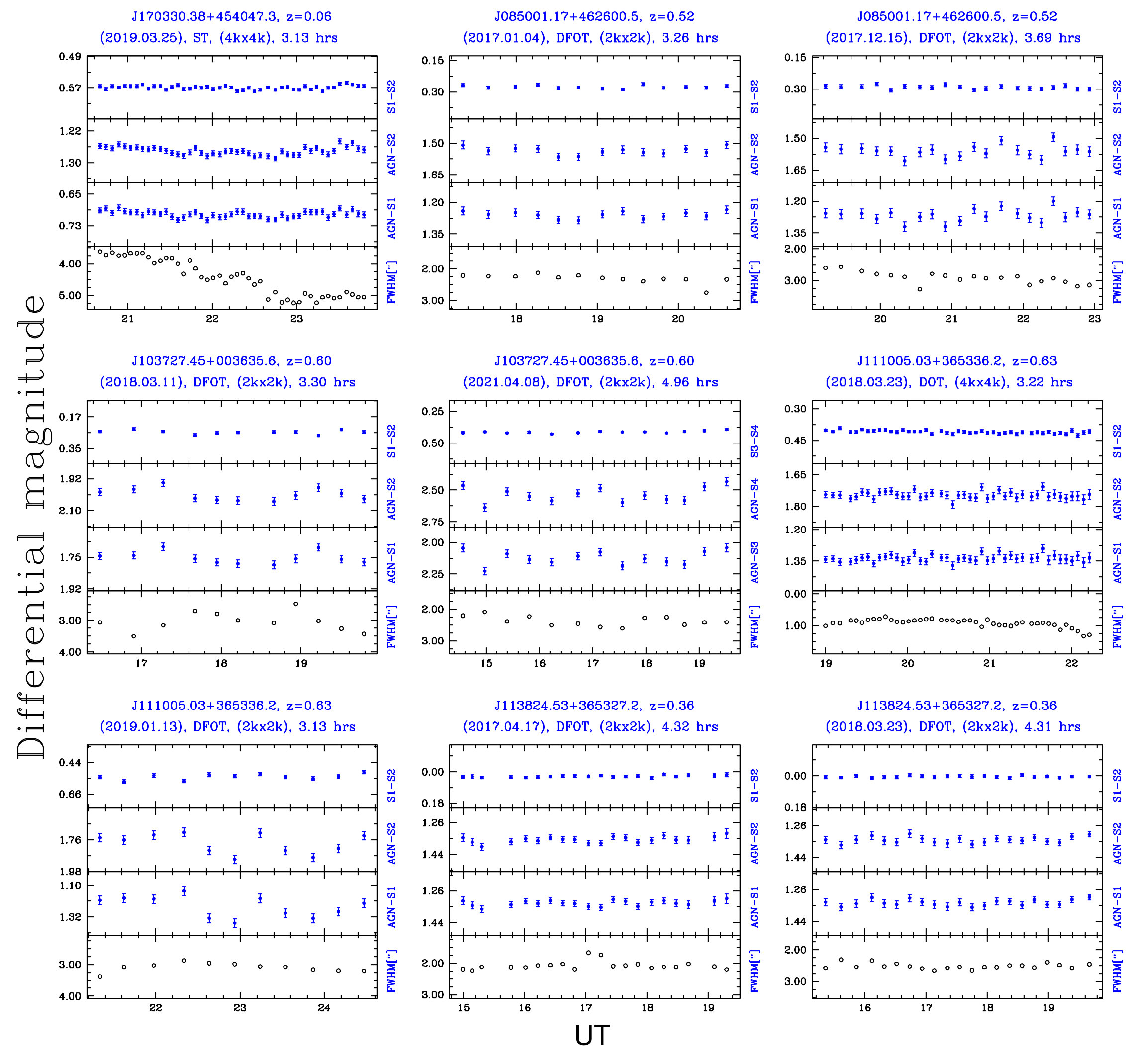}
\caption[]{Similar to Fig.~\ref{fig:lurve 1}, but for the first 4 non-jetted-RLNLSy1s from our sample of 8 non-jetted-RLNLSy1s.}
\label{fig:lurve 5}
\end{figure*}

\begin{figure*}
\ContinuedFloat

\includegraphics[height=21cm, width=19cm]{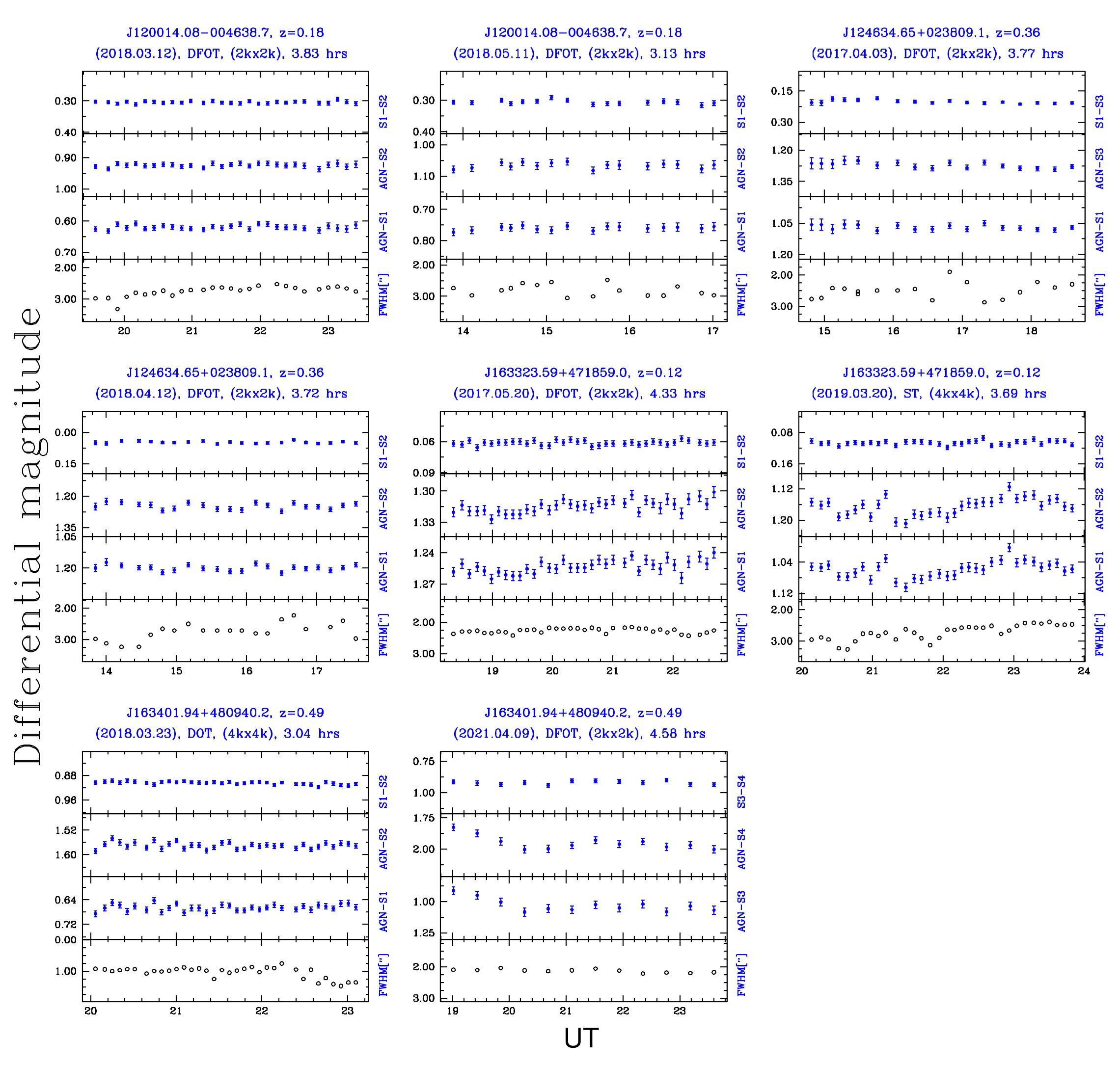}
\caption[]{(Continued) DLC for the last 4 non-jetted-RLNLSy1s from our sample of 8 non-jetted-RLNLSy1s.}
\label{fig:lurve 6}
\end{figure*}

\bibliography{main}
\label{lastpage}

 \end{document}